\documentclass[final,12pt]{elsarticle}

\usepackage{graphicx}
\usepackage{wrapfig}
\usepackage{lscape}
\usepackage{rotating}
\usepackage{epstopdf}
\usepackage{eurosym}
\usepackage{color}
\usepackage{textcomp}
\usepackage{subfigure}
\usepackage{amssymb}

\begin{document}
\begin{frontmatter}
%\frontmatter
%\box{ancora}


%\textcolor[rgb]{1,0,0.3}{Therefore directionality}

\title{{\small \hspace{10cm} LNGS-LOI 48/15} 
NEWS: Nuclear Emulsions for WIMP Search  \hspace{10cm} {\LARGE Letter of Intent}\hspace{10cm} (NEWS Collaboration)\vspace{1cm}  }

\author[2]{A.~Aleksandrov}
\author[11]{A.~Anokhina}
\author[8]{T.~Asada} 
\author[13]{D.~Bender}
\author[10]{I.~Bodnarchuk}
\author[2,6]{A.~Buonaura} 
\author[2]{S.~Buontempo} 
\author[12]{M.~Chernyavskii}
\author[10]{A.~Chukanov} 
\author[2,6]{L.~Consiglio} 
\author[4]{N.~D'Ambrosio} 
\author[2,6]{G.~De Lellis} 
\author[1,5]{M.~De Serio} 
\author[2,6]{A.~Di Crescenzo} 
\author[4]{N.~Di Marco} 
\author[10]{S.~Dmitrievski} 
\author[11]{T.~Dzhatdoev}
\author[1]{R.~A.~Fini} 
\author[8]{S.~Furuya} 
\author[2,6]{G.~Galati} 
\author[2,6]{V.~Gentile} 
\author[12]{S.~Gorbunov}
\author[10]{Y.~Gornushkin} 
\author[13]{A.~M.~Guler}  
\author[8]{H.~Ichiki}
\author[13]{C.~Kamiscioglu}
\author[13]{M.~Kamiscioglu}
\author[8]{T.~Katsuragawa} 
\author[8]{M.~Kimura}
\author[12]{N.~Konovalova}
\author[9]{K.~Kuge} 
\author[2,6]{A.~Lauria} 
\author[3,7]{P.~Loverre} 
\author[8]{S.~Machii} 
\author[11]{A.~Managadze}
\author[3,7]{P.~Monacelli} 
\author[2,6]{M.~C.~Montesi} 
\author[8]{T.~Naka} 
\author[8]{M.~Nakamura}
\author[8]{T.~Nakano}
\author[1,5]{A.~Pastore} 
\author[11]{D.~Podgrudkov}  %MSU
\author[12]{N.~Polukhina}  %Lebedev
\author[14]{F.~Pupilli} %\fnref{14}}
\author[11]{T.~Roganova}    %MSU 
\author[3,7]{G.~Rosa} 
\author[8]{O.~Sato} 
\author[12]{T.~Shchedrina}  %Lebedev 
\author[1,5]{S.~Simone} 
\author[15,16]{C.~Sirignano}
\author[10]{A.~Sotnikov}
\author[12]{N.~Starkov}
\author[2,6]{P.~Strolin}
\author[8]{Y.~Tawara}
\author[2]{V.~Tioukov} 
\author[8]{A.~Umemoto}  
\author[12]{M.~Vladymyrov}
\author[8]{M.~Yoshimoto}
\author[10]{S.~Zemskova} 
%\author{(NEWS Collaboration)}
\address[1]{INFN Sezione di Bari, Bari, Italy}
\address[2]{INFN Sezione di Napoli, Napoli, Italy}
\address[15]{INFN Sezione di Padova, Padova, Italy}
\address[3]{INFN Sezione di Roma, Roma, Italy}
\address[4]{INFN-Laboratori Nazionali del Gran Sasso, Assergi (L'Aquila), Italy }
\address[14]{INFN-Laboratori Nazionali di Frascati, Frascati (Roma), Italy }
\address[5]{Dipartimento di Fisica dell'Universit\`a di Bari, Italy}
\address[6]{Dipartimento di Fisica dell'Universit\`a Federico II di  Napoli, Napoli, Italy}
\address[16]{Dipartimento di Fisica e Astronomia dell'Universit\`a di Padova, Padova, Italy}
\address[7]{Dipartimento di Fisica dell'Universit\`a di Roma, Rome, Italy}
\address[8]{Nagoya University and KM Institute, Nagoya, Japan }
\address[9]{Chiba University, Chiba, Japan }
\address[10]{JINR-Joint Institute for Nuclear Research, Dubna, Russia }
\address[11]{SINP MSU-Skobeltsyn Institute of Nuclear Physics of Moscow State University, Russia}
\address[12]{LPI-Lebedev Physical Institute of the Russian Academy of Sciences, Moscow, Russia}
\address[13]{METU-Middle East Technical University, Ankara, Turkey}

\begin{abstract}

Nowadays there is compelling evidence for the existence of dark matter in the Universe. A general consensus has been expressed on the need for a directional sensitive detector to confirm, with a complementary approach, the candidates found in conventional searches and to finally extend their sensitivity beyond the limit of neutrino-induced background. We propose here the use of a detector based on nuclear emulsions to measure the direction of WIMP-induced nuclear recoils. The production of nuclear emulsion films with nanometric grains is established. Several measurement campaigns have demonstrated the capability of detecting sub-micrometric tracks left by low energy ions in such emulsion films. Innovative analysis technologies with fully automated optical microscopes have made it possible to achieve the track reconstruction for path lengths down to one hundred nanometers and there are good prospects to further exceed this limit. The detector concept we propose foresees the use of a bulk of nuclear emulsion films surrounded by a shield from environmental radioactivity, to be placed on an equatorial telescope in order to cancel out the effect of the Earth rotation, thus keeping the detector at a fixed orientation toward the expected direction of galactic WIMPs. We report the schedule and cost estimate for a one-kilogram mass pilot experiment, aiming at delivering the first results on the time scale of six years.

\end{abstract}

\end{frontmatter}

%\linenumbers

\section{Introduction}\label{sec:Intro}

Compelling evidence for an abundant, non-baryonic, non-luminous (dark) matter component was collected in the last decades \cite{PdG}. Yet, the nature of the dark matter (DM) remains totally unknown, and the quest for an answer ranks as one of the main issues of the experimental particle physics, astrophysics and cosmology. 

Weakly Interacting Massive Particles (WIMPs) \cite{ref1,ref2} are creditable, theoretically appealing DM candidates. If these massive relics of the early universe do exist, they are expected to be gravitationally bound to the baryonic visible matter. A direct search for WIMPs in the mass range from a few GeV/c$^2$ to a few TeV/c$^2$ could be based on the detection of nuclear recoils induced by WIMP elastic scattering. Cross-sections are not expected to exceed those of weak processes. 

\begin{figure}
	\centering
		\includegraphics[width=0.65\linewidth]{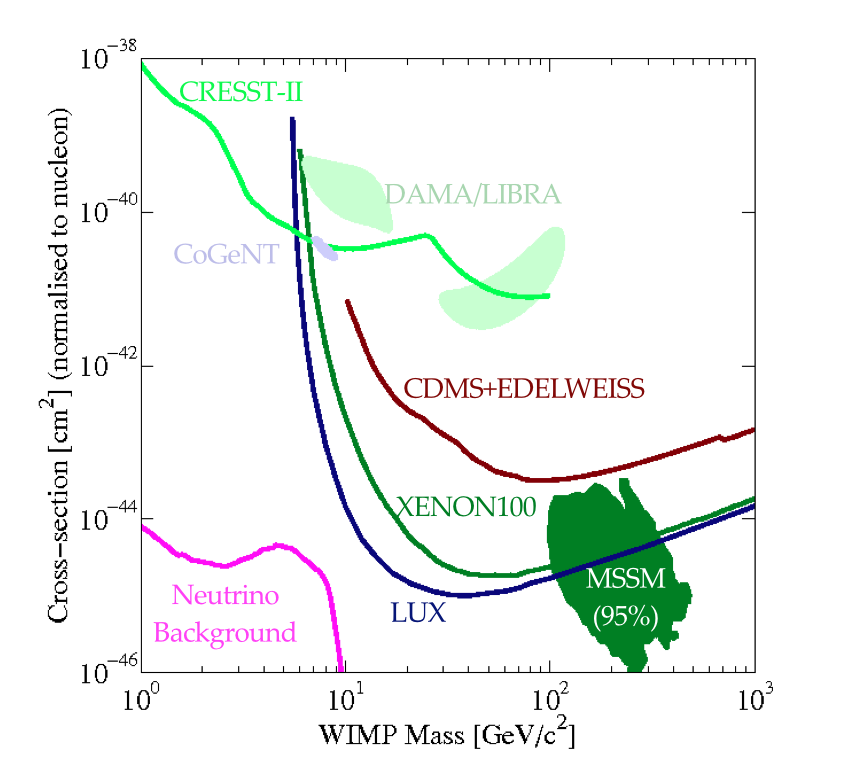} %oltre 0.56 prende l'intera pagina
	\caption{WIMP cross sections (normalized to a single nucleon) for spin-independent couplings versus mass.  
	The DAMA/LIBRA~\cite{DAMA} and CoGeNT~\cite{COGENT} contour regions indicate possible signal events. The 90$\%$ C.L.~upper limits for the CRESST-II~\cite{CRESST}, CDMS+EDELWEISS~\cite{EDELWEISS}, XENON100~\cite{Xenon100} and LUX~\cite{LUX} experiments are shown as solid curves. 
	The green region indicates the predictions from the Minimal Supersymmetrized Standard Model (MSSM) integrated with constraints set by LHC experiments~\cite{MSSM}.
	}
	%\caption{Spin-independent WIMP-nucleon cross-section limits versus WIMP mass. The 90$\%$ C.L. upper limits for the CRESST experiment are shown (solid red, dotted red) together with its contour region reported in phase 1 (light blue) \cite{CRESST}. The limits from Ge-based experiments are shown in green:: SuperCDMS (solid), CDMSlite (dashed) and EDELWEISS (dashed-dotted). The parameter space favoured by CDMS-Si \cite{CDMS-Si} is shown in light green (90 $\%$ C.L.), the one favoured by CoGeNT (99$\%$ C.L. \cite{COGENT}) and DAMA/Libra (3 $\sigma$ C.L. \cite{DAMA-compatibility}) in yellow and orange. The exclusion curves from liquid xenon experiments (90 $\%$ C.L.) are drawn in blue, solid for LUX \cite{LUX}, dashed for XENON100 \cite{Xenon100}.}
	\label{fig:StateOfArt}
\end{figure}

The kinetic energy of scattered nuclei and consequently their range in dense matter would be determined by the WIMP mass and by its velocity relative to a terrestrial target. In the Standard Halo Model the WIMP  speed in the galaxy is supposed to follow a Maxwellian distribution, showing null average values of all the velocity components. The motion of the Solar System through the galaxy, however, creates an apparent wind of dark matter particles, blowing opposite to the direction of the Sun's motion toward the Cygnus constellation. The intensity of this wind, i.e.~the WIMP flux, is expected to be time-modulated due to the Earth motion in the Solar System, with an annual period and a maximum rate in summer \cite{Spergel}. The speed of the Earth in the Solar System is anyway small compared to the speed of the Sun in the Milky Way, so the amplitude of the annual modulation is of the order of a few percent. 
The DAMA experiment \cite{DAMA} at LNGS has indeed reported a signal with a very clear evidence of annual modulation, as a possible indication of DM induced signal. This signal, although statistically extremely significant ($>8$ standard deviations), is controversial because many experiments have already partially or totally  excluded the region allowed by DAMA. Therefore DAMA results remain an intriguing puzzle.
Figure~\ref{fig:StateOfArt} shows the upper limits and contour regions for the WIMP spin-independent 
cross sections, normalized to the scattering on a single nucleon, as function of the WIMP mass. The constraints from SUSY models with the inclusion of LHC results are also shown.
The figure was made with the \texttt{dmtools} web page~\cite{dmtool}.
On the other hand, the angular distribution of the scattered nuclei is peaked around the direction of the apparent dark matter wind. The directional modulation is expected to be stronger than the annual modulation, with a rate of forward-scattered nuclei one order of magnitude higher than the backward-scattered nuclei. Since background sources are expected to be isotropic, the detection of a signal with a preferred direction would provide a powerful discrimination. Directional experiments intend to exploit this effect by measuring the direction of nuclear recoils, and hence the WIMP direction.

In the above sketched WIMP scenario, the key points for the design of an experiment searching for DM with a directional approach are the expected event rate and the expected angular and energy distribution of recoiling nuclei. The expected event rate does not exceed 1 event/kg/year. Such extremely low rates require strong background suppression. The WIMPs mean velocity inside our galaxy is a few hundred kilometers per second at the location of our Solar System. For these velocities, WIMPs interact with ordinary matter mainly via elastic scattering on nuclei. With expected WIMP masses in the range from 10 GeV to 10 TeV, typical nuclear recoil energies are of the order of 1 $\div$ 100 keV. The expected nuclear recoil energy spectrum decreases almost exponentially with energy. To exploit directionality with light-medium mass scattered nuclei, the required spatial accuracy is in the sub-mm domain for gaseous detectors and in the sub-$\mu$m range for solid detectors.  In the first case the low event rate sets the requirement of very large volumes while in the second case an extremely high resolution is required in order to cope with the very short range of the recoil nuclei.

Experiments for dark matter searches based on solid or liquid targets are not able to measure the direction of nuclear recoils. They search for a WIMP signal as an excess of events over the expected background with possibly an annual modulation of the event rate, if sensitive enough. The gaseous detectors, on the other hand, are capable of reconstructing the three-dimensional tracks of nuclear recoils, but their mass and the corresponding sensitivity are rather limited. Current gas-based detectors as DRIFT \cite{DRIFT}, NEWAGE \cite{NEWAGE}, DMTPC \cite{DMTPC} and MIMAC \cite{MIMAC} make use of low-pressure CF$_4$ with a fiducial volume ranging from 3 to 140 g \cite{DRIFT}, thus providing limits only on the spin-dependent WIMP-proton cross-section. 

The use of a solid target would allow to explore lower cross sections in the phase space indicated by recent limits drawn by direct search experiments, the challenge being the shorter track length, $O$(100nm), resulting in the WIMP-nucleus scattering.

The Nuclear Emulsions for WIMP Search (NEWS) project presented here aims at the direct detection of dark matter candidates by measuring the direction of WIMP-induced nuclear recoils. For this challenge, the detector exploits new generation nuclear emulsions with nanometric grains. An R$\&$D conducted by the Nagoya University in collaboration with the Fujifilm Company has established the production of films with nanometric grains for an ultra-high spatial resolution. We do report the results of this R$\&$D and the corresponding development of new fully automated scanning systems capable of detecting such short tracks, with improved optical technologies overcoming the diffraction limit of conventional systems.  We have studied the detection efficiency of nanometric tracks, using ion implantation systems to reproduce nuclear tracks of the same length as expected from WIMP-induced nuclear recoils. A paragraph of this document is devoted to the measurements performed on the neutron yield from intrinsic film radioactivity and more in general to the discussion of potential background sources. Given that nuclear emulsions are time insensitive, the detector will be placed on a standard equatorial telescope to keep its orientation fixed toward the Cygnus constellation. The choice of appropriate shielding materials and  detector layout is also discussed.

Finally we propose the design and construction of a one-kilogram detector for a pilot experiment, acting as a demonstrator of the technology and aiming at scaling it up to a larger scale experiment. The construction, run and data analysis are planned on the time scale of about six years. This experiment will demonstrate the potentiality of the technique and will start constraining the parameter space outlined by the DAMA experiment.

%Expected WIMP flux and angle distribution

\section{NIT: Nano Imaging Tracker} \label{sec:NIT}

After decades of remarkable experimental applications, nuclear emulsions still mantain their attraction as ionizing particle detectors of unmatched spatial and angular resolution.  The first application of fully automated scanning systems to large-scale experiment was for the CHORUS experiment~\cite{CHORUS}. Impressive achievements with new generation systems, more than one order of magnitude faster, allowed the design of the OPERA experiment~\cite{OPERA} while current developments of the technology still inspire the design of high-statistics neutrino experiment with large active target~\cite{SHIP}.

Nuclear emulsions are made of silver halide crystals embedded in a gelatine matrix. When light falls on the emulsion, or ionizing particles pass through it, some of the halide crystals are modified in such a way that  they are turned into grains of silver when 
the emulsion is immersed in a reducing bath (the so-called \emph{developer}). The modifications in the grains caused by the action of light or radiation are  invisible and the effect is referred to as \emph{formation of latent image}. After development, a silver halide emulsion is placed in a second bath, called \emph{fixer}, which dissolves the unaffected grains of silver halide but leaves the small black granules of silver. Finally, the plate is washed and dried~\cite{Emulsion1,Emulsion2,Chap2HandbookOf}. The primary function of the gelatine is to provide a three-dimensional framework which serves to locate the small crystals of the halide and to prevent them migrating during development and fixation. The three-dimensional trajectory of passing through particles can be reconstructed with an optical microscope by connecting all the silver grains produced after development.

The size of silver halide crystals in standard emulsion ranges from 0.1 $\mu$m to 1 $\mu$m. The sensitivity of the emulsion strongly depends on the size of the crystals: the larger the grain, the higher the emulsion sensitivity to ionising radiation. Due to the low recoil energy of a WIMP-scattered nucleus, the expected track length is of the order of a few hundred nanometers. State-of-the-art emulsions produced by the Fuji Co.~\cite{OPERAemulsion} for the OPERA experiment, with a linear dimension of the crystals of 200 nm, are therefore not suitable for Dark Matter searches.

\begin{figure}[tbph]
	\centering\includegraphics[width=1.0\linewidth]{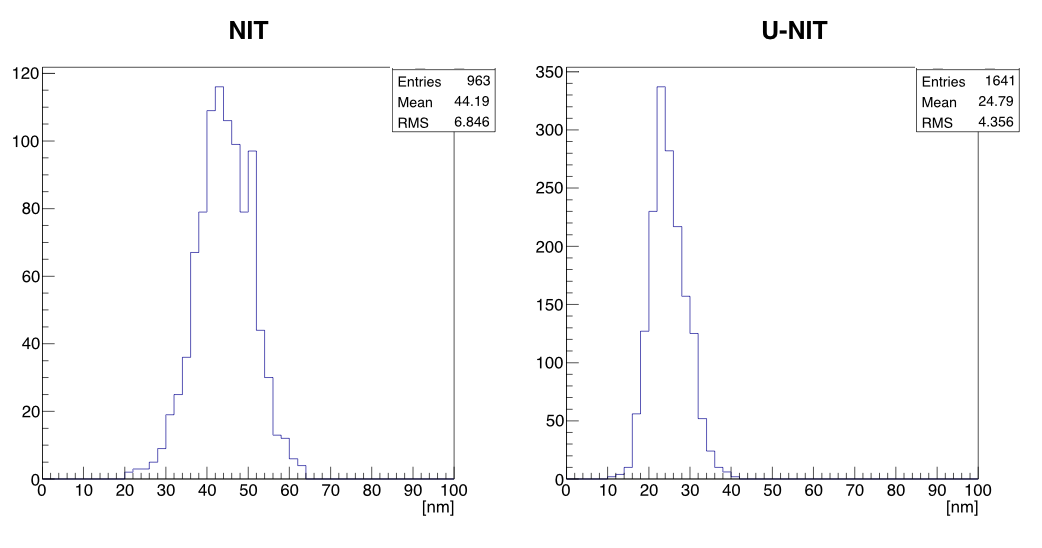}
	\caption{Distribution of the crystal diameter measured with an electron microscope for NIT (left) and U-NIT (right) emulsions. The measurements refer to three different batches.} \label{fig:grains}
\end{figure}

The R$\&$D performed at Nagoya University, in collaboration with Fuji Co. experts, led to the production of novel emulsion films with grain diameters down to a few tens of nm, one order of magnitude smaller than conventional ones. The so-called Nano Imaging Trackers (NIT) and Ultra-Nano Imaging Trackers (U-NIT), have grains of 44.2 and 24.8 nm diameter respectively (see Figure~\ref{fig:grains}). NIT films have a linear density of crystals of about 11 crystals/$\mu$m~\cite{NIT} while U-NIT show 29 crystals/$\mu$m~\cite{U-NIT}. They make  the reconstruction of trajectories with path lengths shorter than 100 nm possible, if analyzed by means of microscopes with enough resolution. \\
%The stability of NIT production was tested by measuring the crystal size in several samples. An average value of 44 nm was obtained, with deviations smaller than 5 nm.

\begin{figure}[tbph]
	\centering\includegraphics[width=0.8\linewidth]{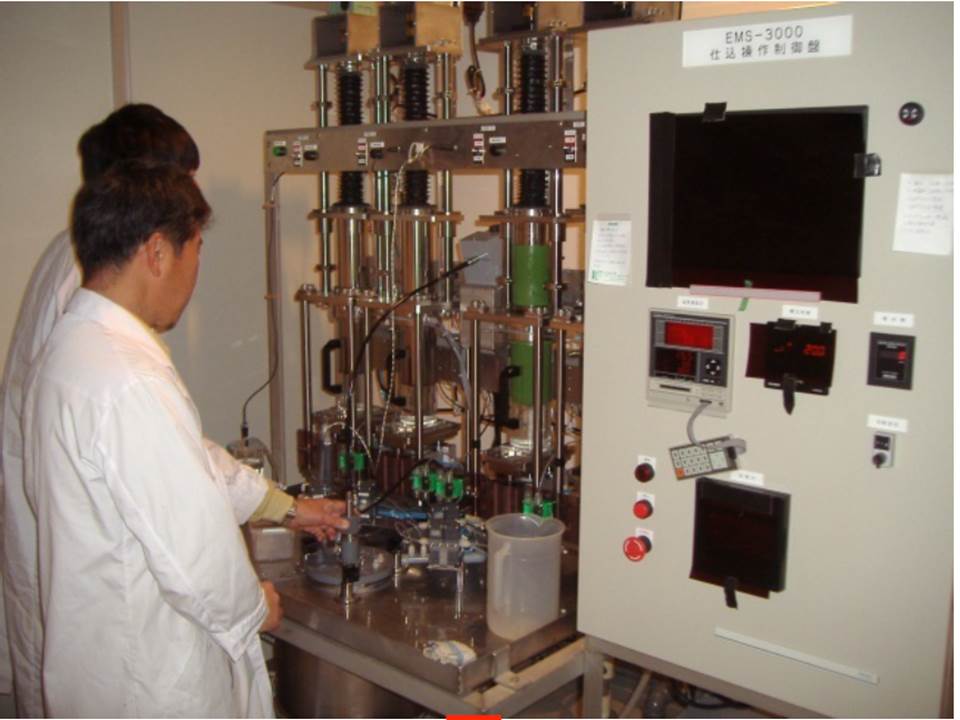}
	\caption{NIT gel production machine.} \label{fig:production-machine}
\end{figure}

NIT are produced in three steps using a dedicated machine (see Figure~\ref{fig:production-machine}). First, the AgBr crystal growth is obtained by mixing in a thermostatic bath AgNO$_3$ and NaBr exploiting the following reaction: 
\begin{equation}
	   \mbox{AgNO}_3 + \mbox{NaBr} \rightarrow \mbox{AgBr} + \mbox{Na}^+ + \mbox{NO}_3^-
\end{equation}
Polyvinyl alcol (PVA) is then added to ensure the uniformity of the grain size of the crystals. NaI, with a concentration of 4\% mol, is also used in order to increase the quantum efficiency in the activation of the crystals. Next, in the desalination phase, AgBr crystals are mixed with the gelatin while the residual extra ions (Na$^+$,NO$_3^-$) are extracted by means of a reduction process. 
A homogeneous crystal distribution is obtained with a centrifugation process at 1000 rpm and $50^\circ$ C. 

\begin{figure}[tbph]
	\centering\includegraphics[width=1.0\linewidth]{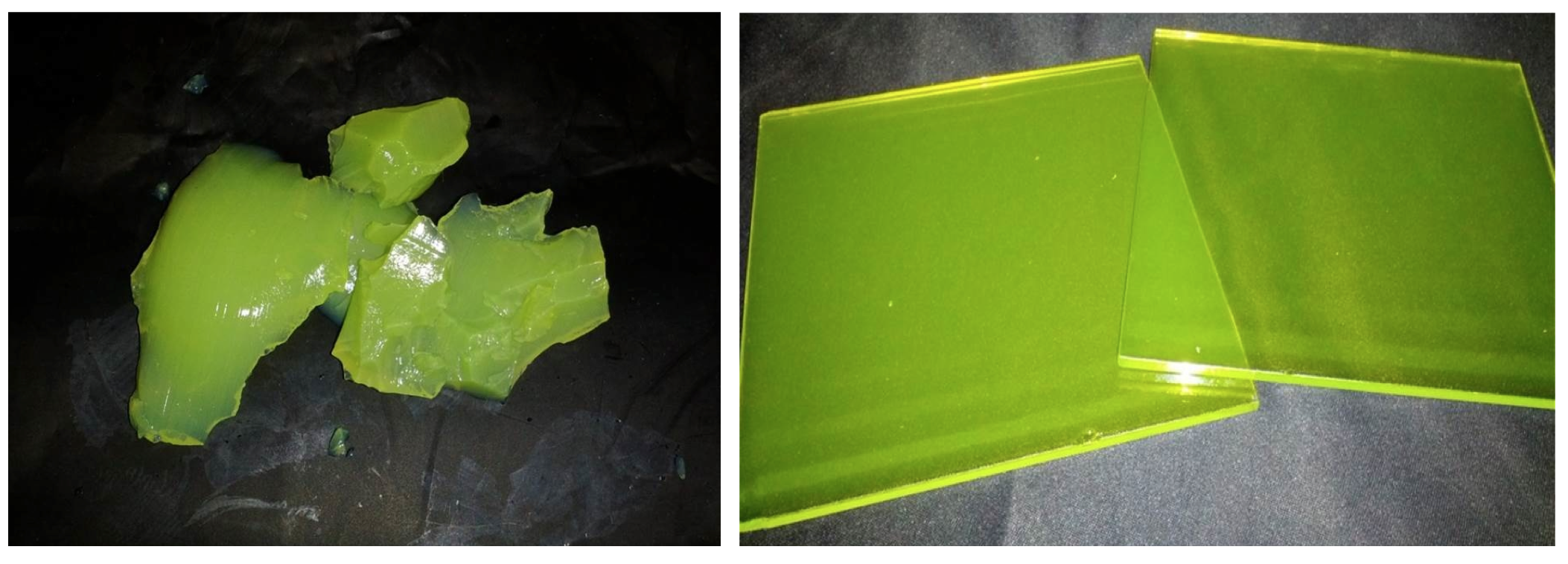}
	\caption{Left: emulsion gel. Right: emulsion gel poured on a glass support.} \label{fig:gel}
\end{figure}

Finally, the emulsion gel obtained with this procedure (see Figure~\ref{fig:gel}, left) is mixed with ultra-pure water and poured on a rigid support (usually plastic or glass) as shown in the right picture of Figure~\ref{fig:gel}. The production machine is able to produce up to 3 kg of NIT emulsion gel per week. 

The mass fractions of NIT constituents and the chemical composition of NIT emulsions are reported in Tables~\ref{tab:composition} and~\ref{tab:constituents}, respectively. The emulsion composition has been carefully determined for light elements by an elemental analyser (YANACO MT-6) with an  uncertainty of 0.3 \%. The mass fraction of silver and bromine has been measured by an energy dispersive X-ray analysis with an uncertainty of 2\%. The density amounts to 3.43 g/cm$^{3}$.

\begin{table}[htpb]
\centering
\begin{tabular}{c|c}
\hline
Constituent & Mass Fraction \\
\hline
AgBr-I      & 0.78  \\
Gelatin     & 0.17 \\
PVA         & 0.05 \\
\hline
\end{tabular} 
\caption{Constituents of NIT emulsions} \label{tab:composition}
\end{table}

\begin{table}[htpb]
\centering
\begin{tabular}{c|c|c}
\hline
Element & Mass Fraction & Atomic Fraction \\
\hline
Ag      & 0.44        & 0.10 \\
Br      & 0.32        & 0.10 \\
I       & 0.019       & 0.004 \\
C       & 0.101        & 0.214 \\
O       & 0.074        & 0.118 \\
N       & 0.027        & 0.049 \\
H       & 0.016         & 0.410 \\
S       & 0.003        & 0.003 \\
\hline
\end{tabular}
\caption{Elemental composition of NIT emulsions.} \label{tab:constituents}
\end{table}

During the whole lifetime of the emulsion and before the development, due to thermal excitation, sensitive crystals can be randomly activated thus resulting in the production of random dark grains: the so-called \emph{fog} (of the order of $1 \div 10 \slash(10 \mu$m$)^3$ for OPERA emulsions) represents a potentially dangerous source of background when looking for very short track length ($O$(100nm)) made of only two consecutive dark grains. In this case, indeed, if the fog density is too high, the probability that two fog grains are close enough to mimic a signal track is not negligible. A recent R$\&$D led to a new chemical development procedure resulting in a suppression of the fog density of one order of magnitude: using a low-temperature ($5^\circ$C) developer based on MAA-1 a fog density of $\sim0.1 \slash(10 \mu$m$)^3$ has been achieved. Moreover, 
fog grains show a rather different contrast and shape with respect to radiation sensitized grains. These important features can be exploited to enhance the signal to background ratio, as it will be explained in Section~\ref{sec:read-out}. %The list of chemicals used for NIT development is reported in Table \ref{tab:chemicals}.

%\begin{table} [htb]
%	\centering
%		\begin{tabular} {c} 
%		\hline
%		List of chemicals \\
%		\hline
%			Na$_2$SO$_3$ \\
%			(L+)-Ascorbic  Acid \\
%			p-(Methylamino)phenol Sulfate (HOC$_6$H$_4$NHCH$_3$)$_2$ $\cdot$ H$_2$SO$_4$ \\
%			Sodium Metaborate Tetrahydrate \\
%			(NaBO$_2$ $\cdot$ 4H$_2$O)\\
%			KBr \\
%			NaOH (1N) \\
%		\hline
%		\end{tabular}
%	\caption{Chemicals needed for emulsion MAA-1 based developement.}
%	\label{tab:chemicals}
%\end{table}

\section{Experimental concept} \label{sec:expConcept}

NEWS is a very innovative approach for a high sensitivity experiment aiming at the directional detection of WIMPs: the detector is based on recent developments of the nuclear emulsions technology allowing to reach an extremely high spatial resolution. 

\begin{figure}[htbp]
	\centering
		\includegraphics[width=0.6\linewidth]{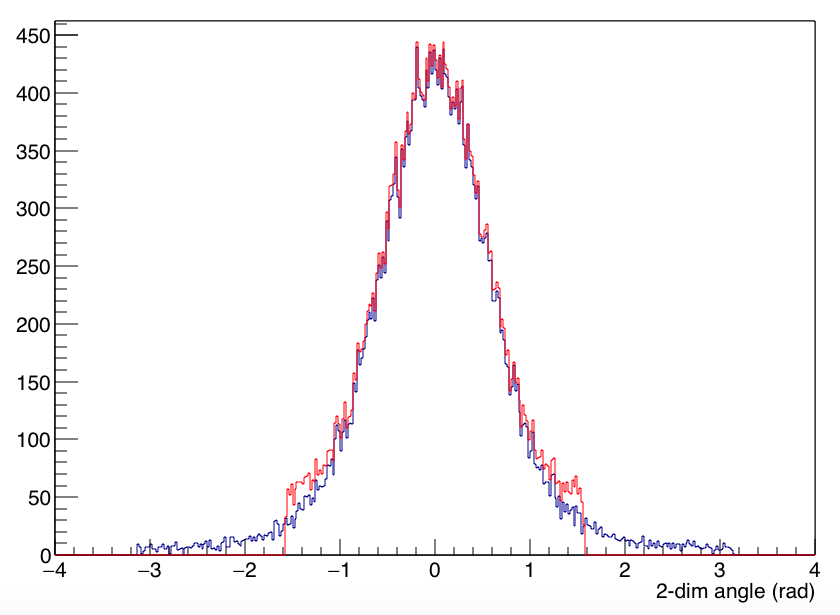}
	\caption{WIMP 2-dim angle distribution on a plane containing the average WIMP wind direction (blue curve).
	The red curve shows the same angle if one is not sensitive to the forward/backward direction.}
	\label{fig:wimp_direction}
\end{figure}

The detector is conceived as a bulk of nuclear emulsions acting both as a target and as a tracking device surrounded by a shield (see Section~\ref{sec:set-up}) to reduce the external background. The detector will be placed on an equatorial telescope in order to absorb the earth rotation, thus keeping the orientation towards the Cygnus constellation  fixed. The emulsion films will lie with their surface permanently parallel to the expected average WIMP wind direction. Figure~\ref{fig:wimp_direction} shows the distribution of the WIMP incoming angle, in the laboratory  frame, projected on a plane containing the average WIMP wind direction. 
The majority of WIMPs are directed forward with a peak at zero.
The superimposed red curve shows the same angle if one is not sensitive to the forward/backward direction.
 The angular distribution of the trajectories of WIMP-scattered nuclei is therefore expected to be anisotropic.
%, being the average track angle peaked in the direction of the apparent WIMP wind in the laboratory frame. 

\begin{figure}[tbph]
	\centering
		\includegraphics[width=0.6\linewidth]{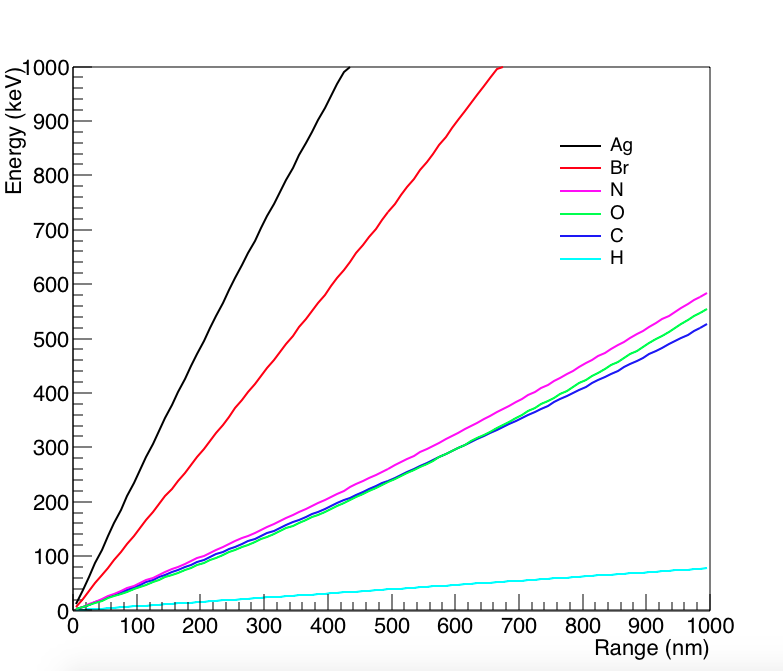}
	\caption{Correlation between the  track length of the recoiled nuclei and their kinetic energy, for different target nuclei in  NIT emulsions.}
	\label{fig:correlation}
\end{figure}

The presence in the emulsion gel of lighter nuclei such as carbon, oxygen and nitrogen, in addition to the heavier nuclei of silver and bromine, is a key feature of the NEWS project, resulting in a good sensitivity to WIMPs with both light and heavy masses. The sensitivity indeed strongly depends on the minimum detectable track length. The path length of the recoiled track depends in turn on the kinetic energy of the scattered nucleus, being the kinematics determined both by the mass of the incident WIMP and by that of the target nucleus.  The correlation between the track length of the recoiled nucleus and its kinetic energy is shown in Figure~\ref{fig:correlation} for the different target nuclei. WIMP with a mass of about 100 GeV/c$^2$ prefers Ag and Br as target, producing e.g.~Br recoils with an average kinetic energy of about 50 keV. Although Ag and Br are the most effective targets for WIMP masses in this range, the detection capability is reduced since their ranges are shorter than lighter elements at the same energy. Instead, for a WIMP with a mass around 10 GeV/c$^2$, the kinematics favours lighter nuclei that, for a given kinetic energy, have a longer range. Therefore, the contribution of the C, N and O ions is essential for WIMP masses around 10 GeV/c$^2$.

The estimated WIMP rates are of the order of 1 event$\slash$kg$\slash$year, much lower than the usual radioactive backgrounds. For this reason, the detector has to be placed underground to be protected from cosmic-ray induced background. Moreover, a careful control of the radioactive contamination of the
materials used for the detector construction and a precise estimation of
the corresponding induced background are needed. We will discuss in detail
the most relevant background sources for the WIMP search with an emulsion
based detector on the mass scale of a few kilograms.

After the exposure, the emulsion films composing the target will be developed and the whole detector volume will be analyzed by using fully automated scanning systems. The read-out (see Section~\ref{sec:read-out}) is performed in two phases. In the first phase a fast scanning is performed (see Section \ref{sec:optical-read-out}) by means of an improved  version of the optical microscope used for the scanning of the OPERA films (\cite{ESS,S-UTS}). By this step a fast pre-selection of the candidate signal tracks with a relatively low spatial resolution (200 nm) can be achieved. In order to resolve the nanometric grains belonging to a signal tracks and to enhance the signal to background ratio, a further scanning of the pre-selected tracks with a ultra-high resolution  scanning system is foreseen (see Section~\ref{sec:plasmon}). The final resolution for the reconstruction of nuclear recoil tracks is estimated to be between $10$ and $20$ nm in position and better than $15 ^\circ$ in angle. 
 
\section{Read-out technique} \label{sec:read-out}

In the NEWS experiment the expected WIMP signal will consist of short-path, anisotropically distributed, nuclear recoils over an isotropically distributed background. 
The search for signal candidates requires the scanning of the whole emulsion volume. 

The read-out system has therefore to fulfill two main requirements: a fast, completely automated, scanning system is needed to analyse the target volume 
over a time scale comparable with the exposure; the spatial resolution has to be improved by more than one order of magnitude compared to that achieved with standard emulsion films, reaching the challenging value of a few tens of nanometers, in order to ensure high efficiency and  purity in the selection of signal candidates.

The analysis of NIT emulsions is performed with a two-step approach: a fast scanning with a state-of-the-art resolution for the signal preselection followed by a pin-point check of preselected candidates with unprecedented nanometric resolution to further enhance the signal to noise ratio and perform very accurate measurements of the range and the recoil direction. These two steps are discussed in the next sub-sections.

\subsection{Optical microscopy for candidate selection} \label{sec:optical-read-out}

\begin{figure}[htbp]
	\centering%
		\subfigure[Japanese prototype\label{fig:mic_Nagoya}]%
			{\includegraphics[width=0.5\linewidth]{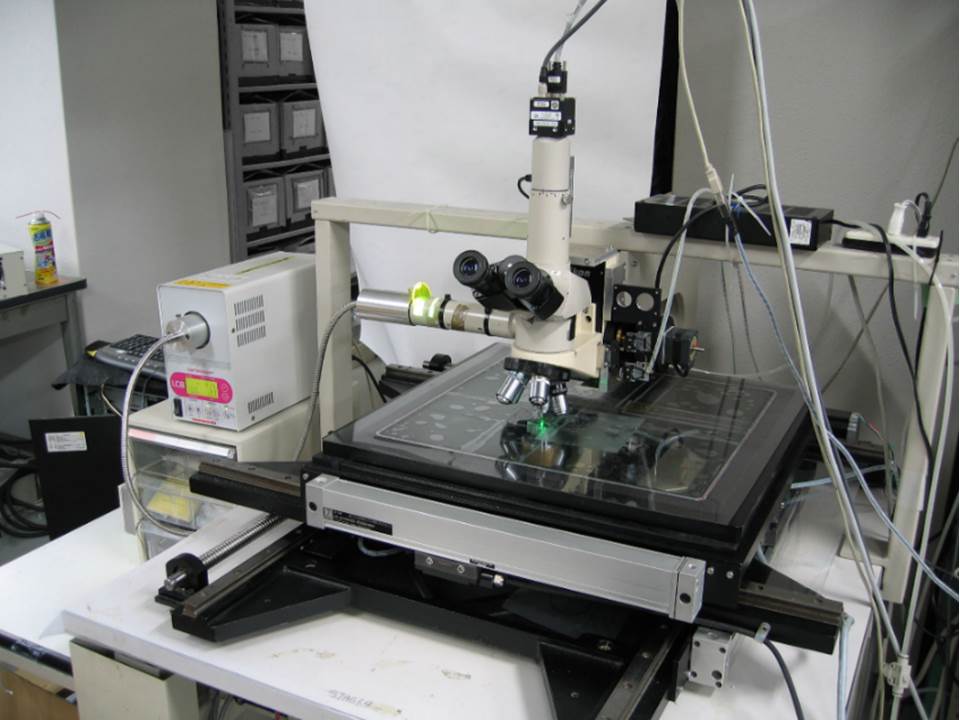}}\qquad\qquad
		\subfigure[Prototype developed by INFN groups \label{fig:mic_LNGS}]%
			{\includegraphics[angle=-90,width=0.5\linewidth]{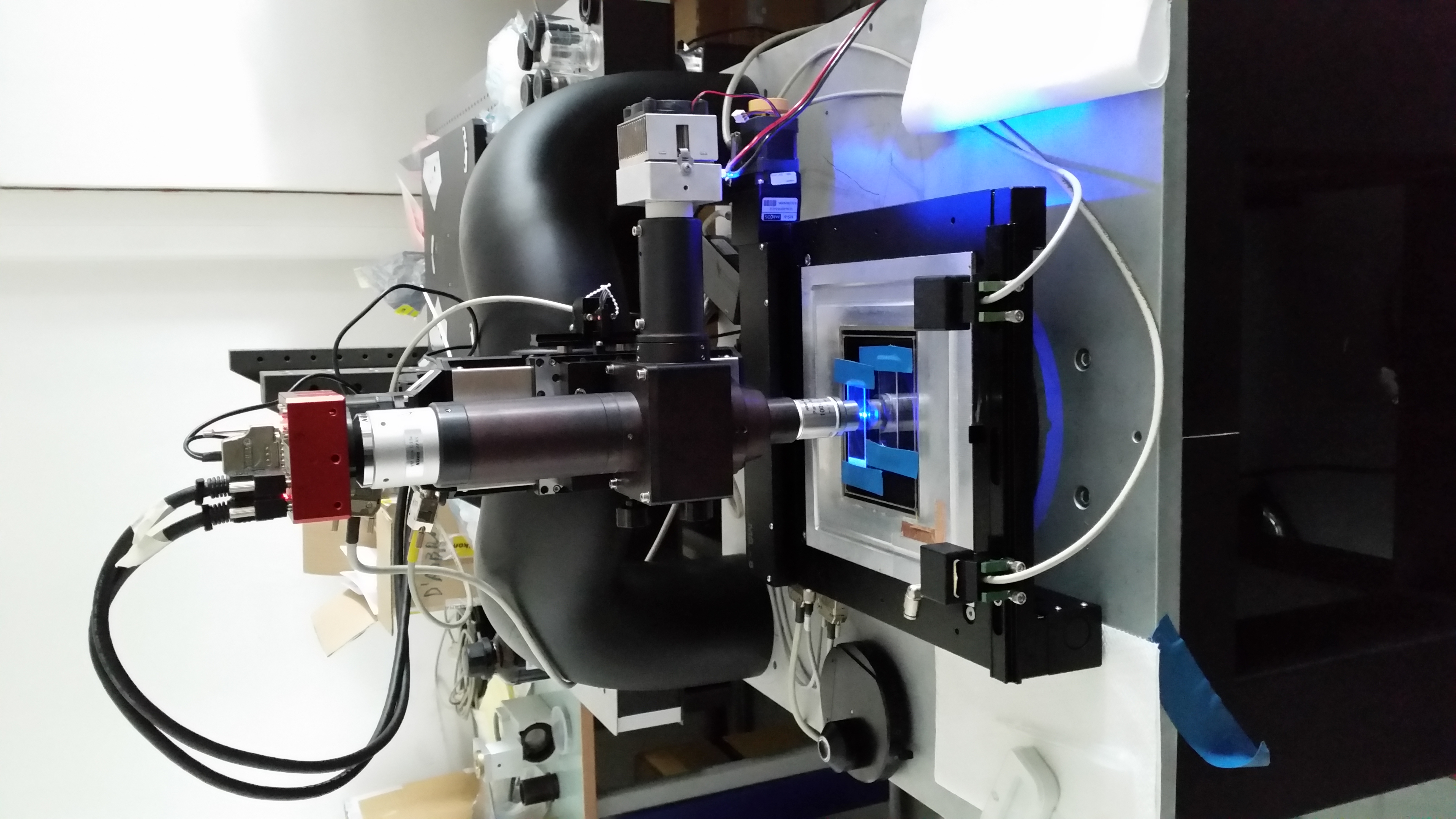}}\qquad\qquad
\caption{Optical scanning systems modified for the analysis of NIT. \ref{fig:mic_Nagoya} Prototype installed at Nagoya University. \ref{fig:mic_LNGS} Prototype installed at LNGS and Naples scanning laboratories. \label{fig:mics}}
\end{figure}

The members of the NEWS Collaboration own state-of-the-art experience of large-scale fast automated scanning  with a spatial resolution of about $1\mu$m and an angular resolution of about 1 mrad, as currently applied in the OPERA experiment \cite{OPERAhowTo}: the European Scanning System (ESS~\cite{ESS}) in  Europe and the Super-Ultra Track Selector (S-UTS~\cite{S-UTS}). 
In the last years an R\&D program aimed at improving the ESS performances was carried by INFN groups, leading to prototypes with resolution improved by one order of magnitude, achieving a speed of almost 200 cm$^2$/h~\cite{ESS-new}.  A new system is being developed in Japan (the Super-Ultra Track Selector), aiming at increasing the scanning speed up to 5000 cm$^2$/h.

Stepping into the nano imaging domain requires substantial upgrades of the OPERA-style scanning systems. New prototypes (see Figure~\ref{fig:mics}) were already set-up both in Japan and in Italy, featuring: 
\begin{itemize} 
	\item higher magnification of the objectives lens, from 50x to 100x
	\item higher numerical aperture, from 0.8 to 1.45
	\item higher optical contrast (illumination by reflected light instead of transmitted light).
	\item light with green or blue wavelength to improve the resolution
	\item high pixel to micron ratio ($\sim$ 28 nm/pixel), one order of magnitude better than the systems used in OPERA
	\item high resolution (4Mpx) and high speed (563 fps) CMOS camera. 
\end{itemize}
In parallel with the hardware improvements, the development of a new acquisition software and a new tracking algorithm has been carried out: the high data rate (1.7 GB/s), a factor 4 higher than the ESS one due to
the improved sensor resolution, has required the use of last generation acquisition boards (Matrox Radient eCL SFCL/DFCL). As a consequence, a more powerful computing system, exploiting a GPU (Graphic Processing Unit) based architecture, has been implemented.

In order to evaluate the performances of the new scanning systems, extensive tests were performed with exposures of NIT to slow ions and neutron beams. Results are discussed here.

The starting point of the emulsion scanning is the image analysis to collect clusters made of dark grains at several depths across the emulsion plate thickness. Given the intrinsic resolution of the optical microscope ($\sim$ 200 nm), the sequence of several grains making a track of a few hundred nanometers, appears as a single cluster. Therefore, the key element to distinguish clusters made of several grains from clusters made of a single grain produced by thermal excitation (fog) is the analysis of their shape. A cluster made of several grains indeed tends to have an elliptical shape with the major axis along the direction of the trajectory, while a cluster produced by a single grain tends to have a spherical shape.

In order to simulate the effect of a WIMP-induced nuclear recoil and to measure the efficiency and the resolution of the new optical prototype, a test beam with low velocity ions was performed. We used both a Kr ion beam with energies of 200 and 400 keV~\cite{ShapeAnalysis} and a C ion beam with energies of 60, 80 and 100 keV. Kr and C ions of such energies produce in emulsion tracks with a length in the range 100$\div$300 nm. These ions were implanted in the emulsions using an implantation facility of low speed ions at Nagoya University. 

\begin{figure}[tbph]
	\centering
		\includegraphics[width=0.6\linewidth]{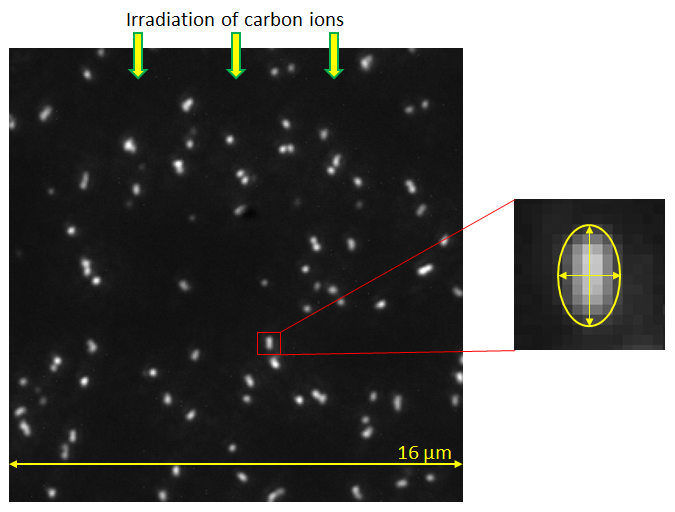}
	\caption{Kr ions implanted on NIT films. The image is taken with an optical microscope. The selection of candidate tracks is based on the elliptic fit of the clusters}
	\label{fig:KrIon-ShapeAnalysis}
\end{figure}

\begin{figure}[tbph]
	\centering
		\includegraphics[width=1.0\linewidth]{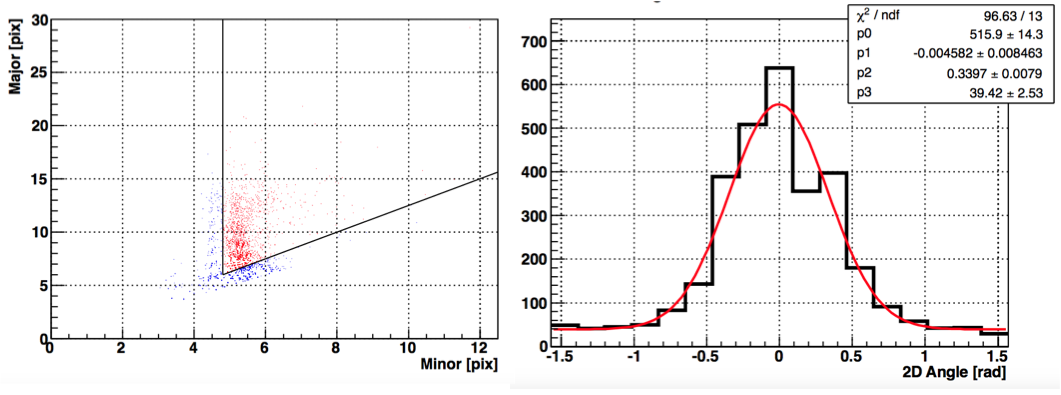}
	\caption{Left: scatter plot of major and minor axes for clusters analysed with an elliptical fit in a 60 keV C ion test beam. Signal tracks are shown as red dots, fog grains in blue.  Right: angular distribution of 60 keV C ion tracks selected by the ellipticity cut.}
	\label{fig:shapeAnalysis60}
\end{figure}

\begin{figure}[tbph]
	\centering
		\includegraphics[width=1.0\linewidth]{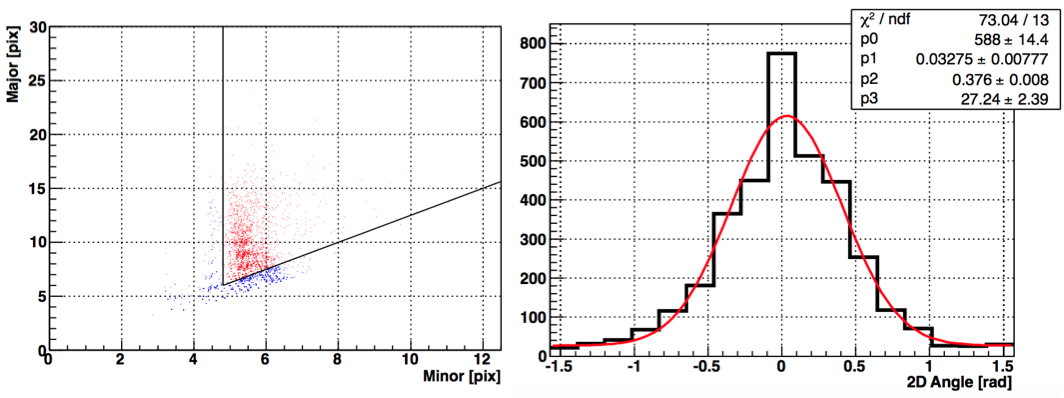}
	\caption{Left: scatter plot of major and minor axes for clusters analysed with an elliptical fit in a 80 keV C ion test beam. Signal tracks are shown as red dots, fog grains in blue.  Right: angular distribution of 80 keV C ion tracks selected by the ellipticity cut.}
	\label{fig:shapeAnalysis80}
\end{figure}

When analysed with the optical microscope, submicrometric tracks produced by Kr and C ions appear as shown in Figure~\ref{fig:KrIon-ShapeAnalysis}. Although silver grains belonging to the tracks are not distinguishable and appear as a single cluster, the elongated form of the cluster is clearly visible~\cite{ShapeAnalysis2}. An elliptical fit of the cluster shape allows a clear separation between fog grains and signal tracks: the latter ones are expected to have ellipticity larger than a given threshold, typically 1.25 or higher (see left plot of Figure~\ref{fig:shapeAnalysis60} and \ref{fig:shapeAnalysis80}). 

The angular distributions of 60 and 80 keV C ions are reported in the right plot of Figure~\ref{fig:shapeAnalysis60} and Figure~\ref{fig:shapeAnalysis80}, respectively. A peak corresponding to the direction of the implanted ions is clearly visible; the width of the distribution corresponds to the angular resolution, amounting to 360 mrad. The angular resolution is given by the convolution of the intrinsic resolution with the angular deviations caused by the scattering in the material. For low energy (below 100 keV) tracks, the scattering cannot be neglected.

\begin{figure}[tbph]
	\centering
			\subfigure[\label{fig:angularResMethod}] {\includegraphics[width=0.48\linewidth]{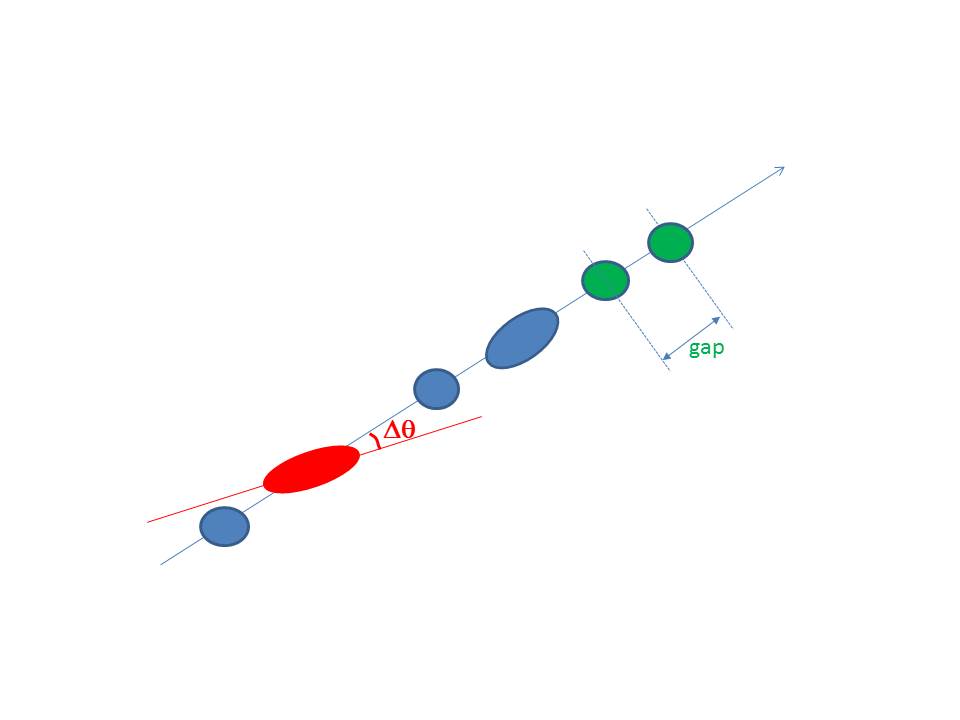}}
			\subfigure[\label{fig:angularResAndrey}] {\includegraphics[width=0.51\linewidth]{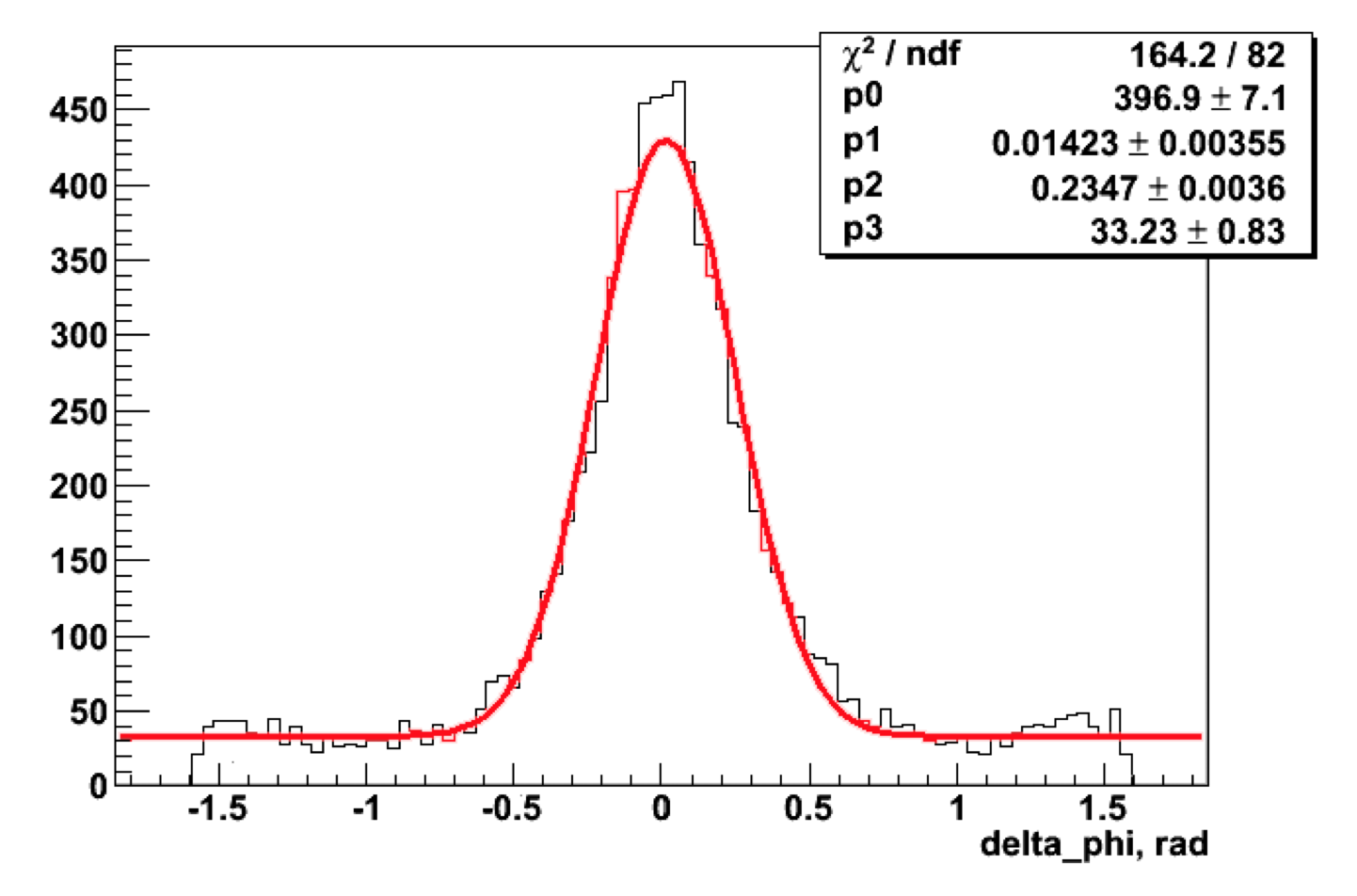}}
	\caption{Left: sketch of the method used for the evaluation of the intrinsic angular resolution. Right: intrinsic angular resolution of the optical scanning system.}
	\label{fig:AngularResolution}
\end{figure}

In order to evaluate the intrinsic angular resolution of the scanning system we analysed an emulsion sample exposed to a 2.8 MeV neutron beam at the Fusion Neutronics Source (FNS) of the Japan Atomic Energy Agency (JAEA).
In this case the track length distribution of neutron-induced proton recoils shows a wider range, up to a few hundred of micrometers. 
A sample of tracks with length of the order of few tens of micrometers and made by a sequence of several elliptical clusters was selected, being the scattering effect negligible for them.
The same ellipticity cut applied in the previous analysis was used for the selection of the clusters.
For each cluster, the angular difference $\Delta \theta$ between its major axis and the fitted track was evaluated (see Figure~\ref{fig:angularResMethod}). The distribution of $\Delta \theta$ shows a gaussian shape, as shown in Figure~\ref{fig:angularResAndrey} with a width corresponding to the intrinsic angular resolution and amounting to 230 mrad. This value represents the intrinsic angular resolution achieved with fully automated scanning systems, by far the best resolution achieved with direction sensitive detectors in this energy range. The simulation shows that this result is compatible with the measurement reported above when the scattering contribution is included.

Tracks selected with the shape analysis were validated using the X-ray microscope~\cite{NakaX-ray}. This technique features a higher resolution (of the order of 60 nm) but a slower scanning speed when compared with the optical microscopy. 
The analysis of a few hundred $\mu$m$^2$ takes about 100 s.
The X-ray microscopy can therefore be used only to check a sample of already selected candidate tracks: X-ray analysis was used to demonstrate the principle of selection by elliptical shape analysis and measure the efficiency achievable with the optical microscopy.

\begin{figure}[tbph]
	\centering
		\includegraphics[width=0.8\linewidth]{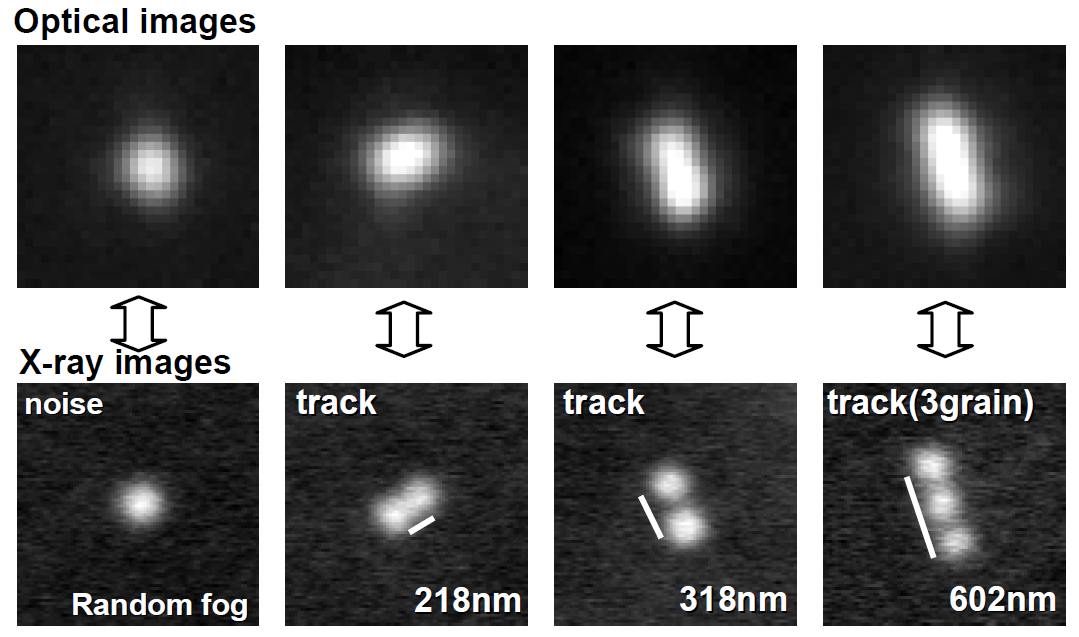}
	\caption{Comparison between reconstructed tracks of a few hundred nanometers length with the optical microscope and with the X-ray microscope.}
	\label{fig:x-ray_confirmation}
\end{figure}

The comparison of optical and X-ray images of candidate tracks is reported in Figure~\ref{fig:x-ray_confirmation}: the high resolution of the X-ray microscope allows to resolve grains belonging to submicrometric tracks thus providing the final discrimination between signal and background. 

\begin{figure}[tbph]
	\centering
		\includegraphics[width=0.6\linewidth]{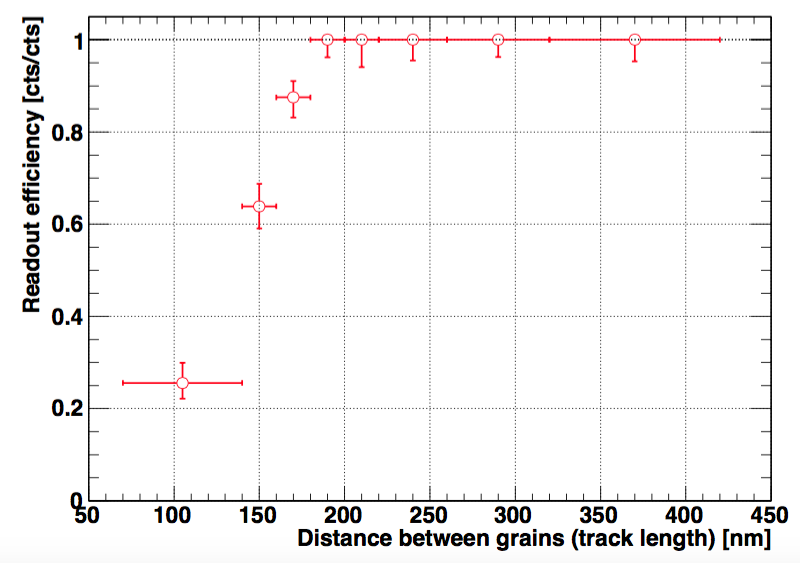}
	\caption{Efficiency of the elliptical fit analysis versus the track length when an ellipticity of 1.25 is used as a threshold. }
	\label{fig:eff_vs_length}
\end{figure}

In Figure~\ref{fig:eff_vs_length} the detection efficiency of the optical system as a function of the track length is shown: the efficiency is obtained first selecting a set of multi-grain tracks with the X-ray microscope and then scanning them with the optical one and applying the shape analysis. In this test an optical microscope with a pixel to micron ratio of 55 nm/pixel was used. Results show that the efficiency reaches 100$\%$ above 180-200 nm.

\begin{figure}[tbph]
	\centering
		\includegraphics[width=0.8\linewidth]{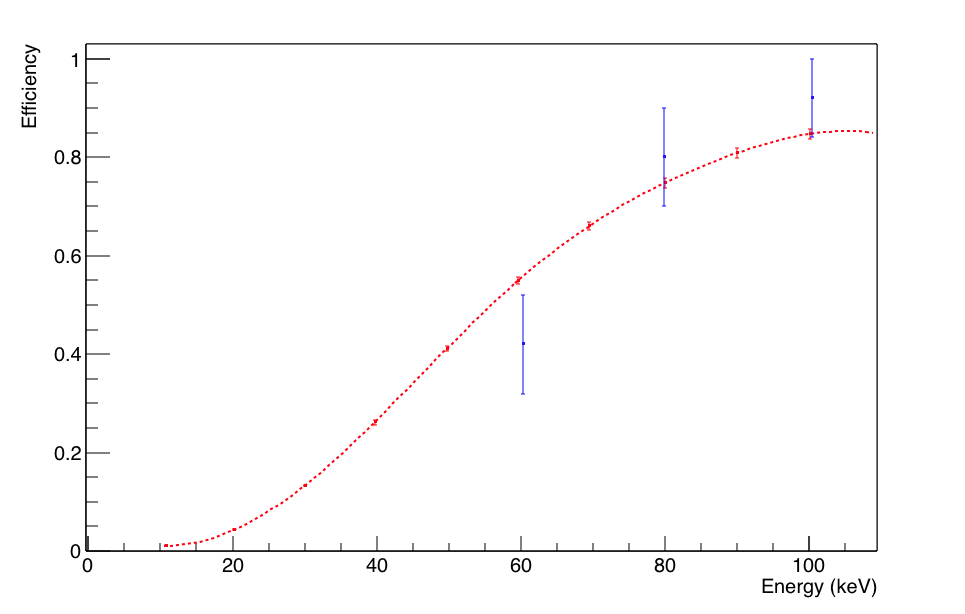}
	\caption{Efficiency of the elliptical fit analysis versus the C ion energy when an ellipticity of 1.25 is used as a threshold.  MC simulation (red line) well describes the data (blue points).}
	\label{fig:eff_vs_energy}
\end{figure}

In Figure~\ref{fig:eff_vs_energy} the efficiency as a function of the recoil energy for C ions of 60, 80 and 100 keV, is shown: MC simulations (red line) well describes the data (blue points). 
%A pixel size of 28 nm was used in this case.
It is worth noting that the capability of reconstructing low energy tracks (E $<$ 40 keV), corresponding to shorter path lengths, although with a lower efficiency, could significantly enhance the sensitivity to low WIMP mass regions.

The scanning speed of the prototype currently used for the shape analysis is of about 25 mm$^2$/h.

\subsection{Beyond the limits of the optical scanning for candidate validation} \label{sec:plasmon}

The use of optical microscopes allows the reconstruction of tracks down to 200 nm. The X-ray microscopy can overcome this limit though being extremely slow if compared with automated optical systems. Being the speed an issue in the analysis of a large mass detector, NEWS aims at the improvement of the spatial resolution enhancing the optical microscopy without using X-ray microscopes.

\begin{figure}[tbph]
	\centering
		\includegraphics[width=1.0\linewidth]{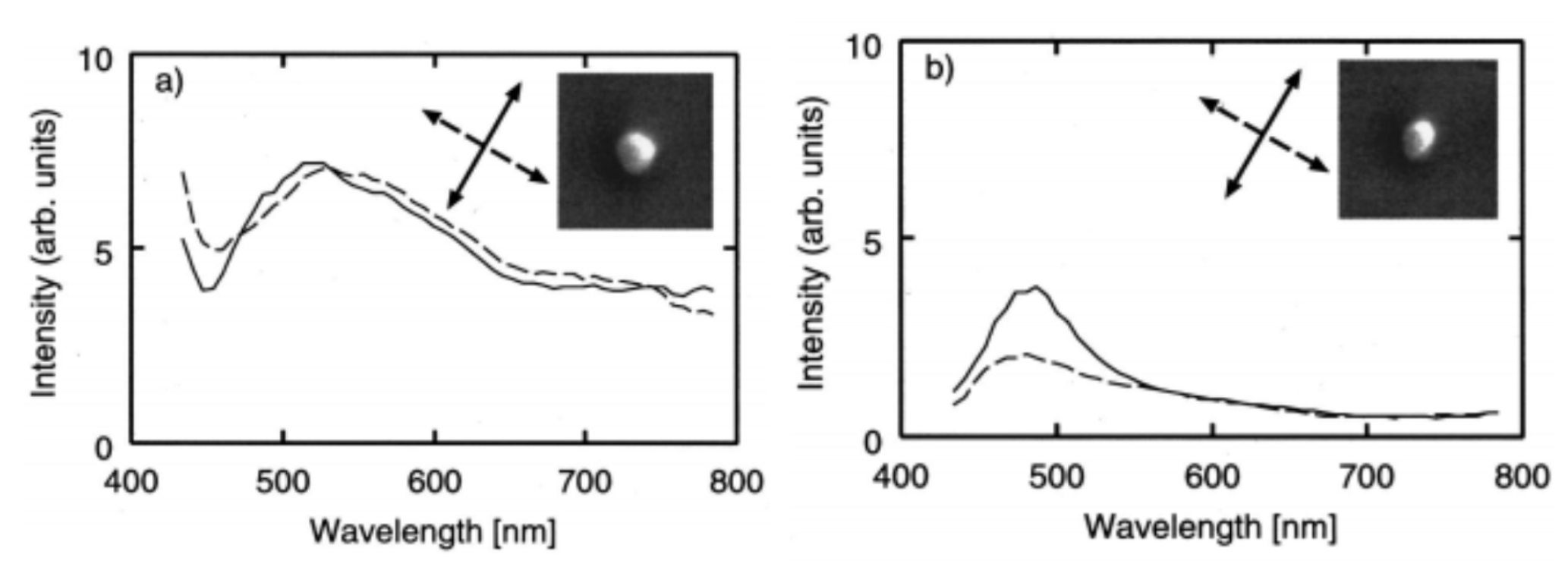}
	\caption{Scattered-light spectra from individual Ag particles with spherical (left) and spheroidal (right) shape \cite{ResonantLightScattering}. The inset shows the 300 $\times$ 300 nm$^2$ SEM image of the particle. Arrows indicate the polarization of the incident light. A dependence of the response on the light polarization is observed for particles with ellipsoidal shape.}
	\label{fig:resonantLight}
\end{figure}

The basic idea is to exploit the resonance effect occurring when nanometric metal grains are dispersed in a dielectric medium~\cite{ResonantLightScattering}. The polarization dependence of the resonance frequencies strongly reflects the shape anisotropy and can be used to infer the presence of non-spherical nanometric silver grains. Figure~\ref{fig:resonantLight} shows the results of the resonant light scattering from individual Ag nanoparticles~\cite{ResonantLightScattering}: spherical particles do not show any different response as a function of the incident polarization, while a deformed sphere is sensitive to the polarization. 

NEWS will use this technology to retrieve track information in NIT emulsions beyond the optical resolution. Images of the same cluster taken with different polarization angles will show a displacement of the position of its barycenter. The analysis of the displacements allows to distinguish clusters made of a single grain from those made of two (or more) grains.

\begin{figure}[tbph]
	\centering
		\includegraphics[width=1.0\linewidth]{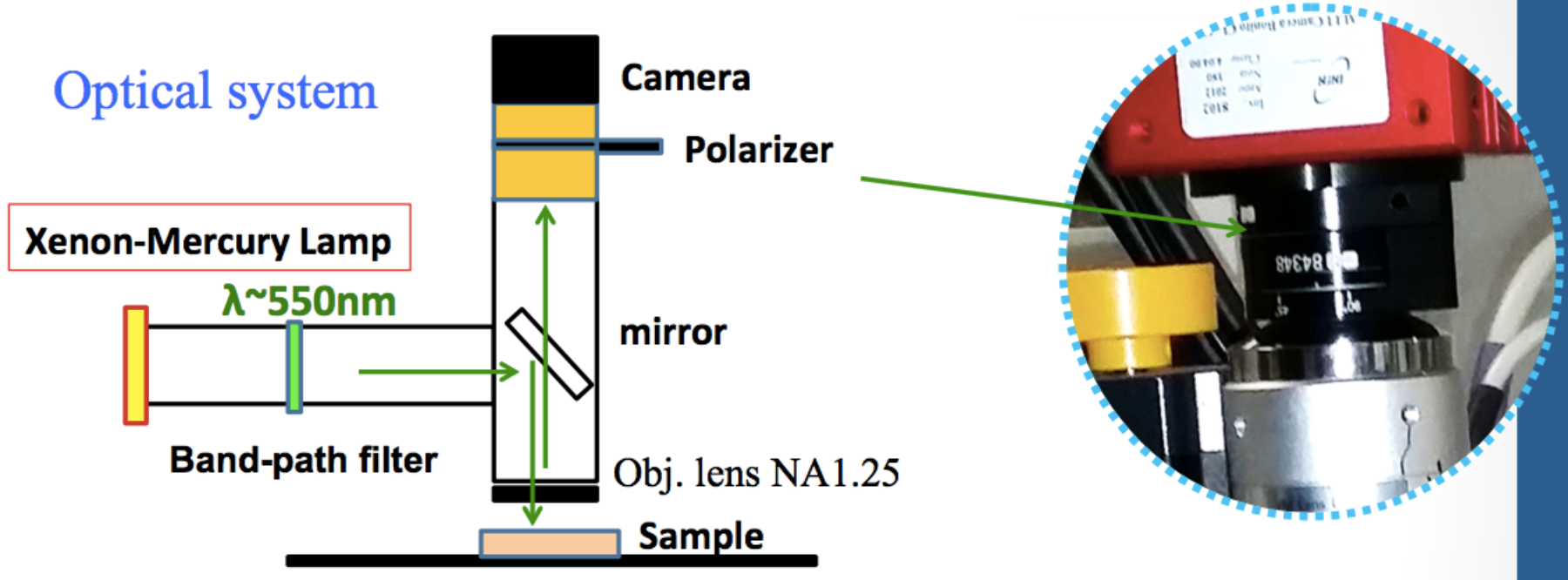}
	\caption{Schematic view of the optical path instrumented with a polarizer to obtain a nanometric resolution with optical microscopes. }
	\label{fig:plasmon_prototype}
\end{figure}

\begin{figure}[tbph]
	\centering
		\includegraphics[width=1.0\linewidth]{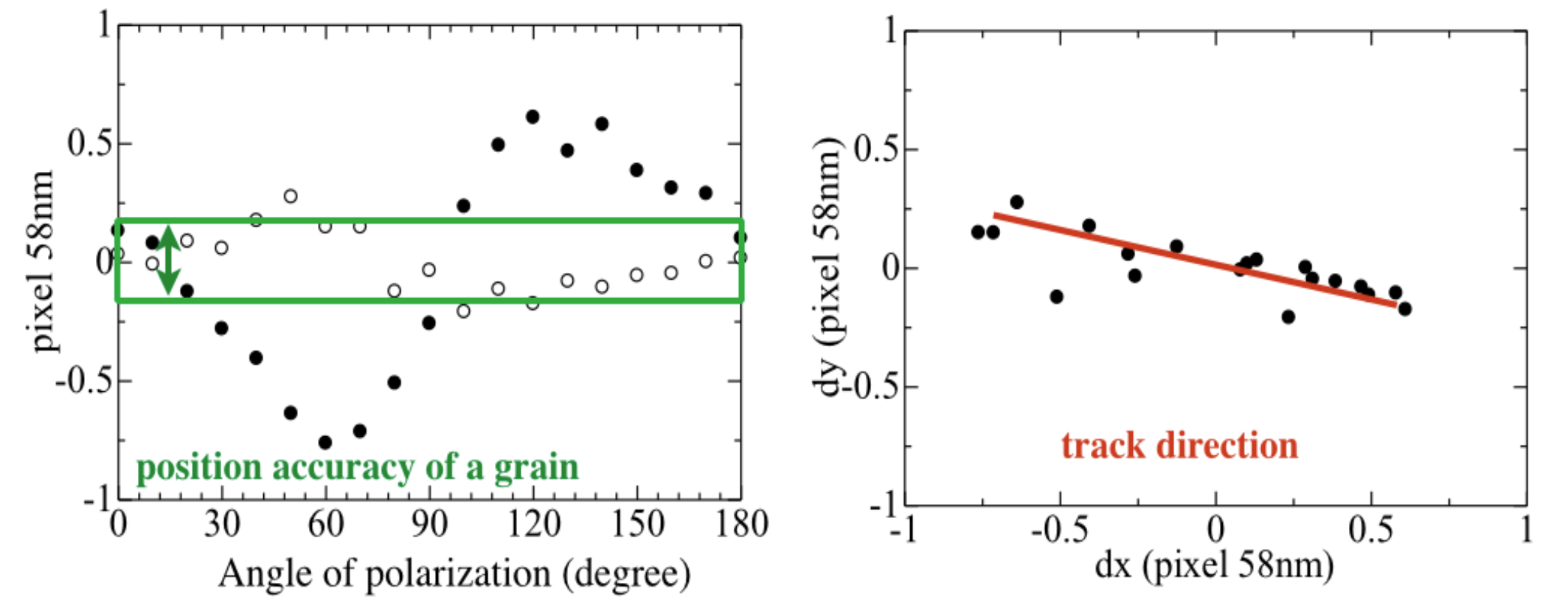}
	\caption{Application of resonant light scattering to an elliptical cluster with ellipticity 1.27. Left plot: $dx$ and $dy$ are the displacements of the cluster barycenter for a given polarization in pixel units (1 pixel = 55 nm). Right plot: track slope fit and its length of about 90 nm.}
	\label{fig:plasmon_analysis1}
\end{figure}

\begin{figure}[tbph]
	\centering
		\includegraphics[width=1.0\linewidth]{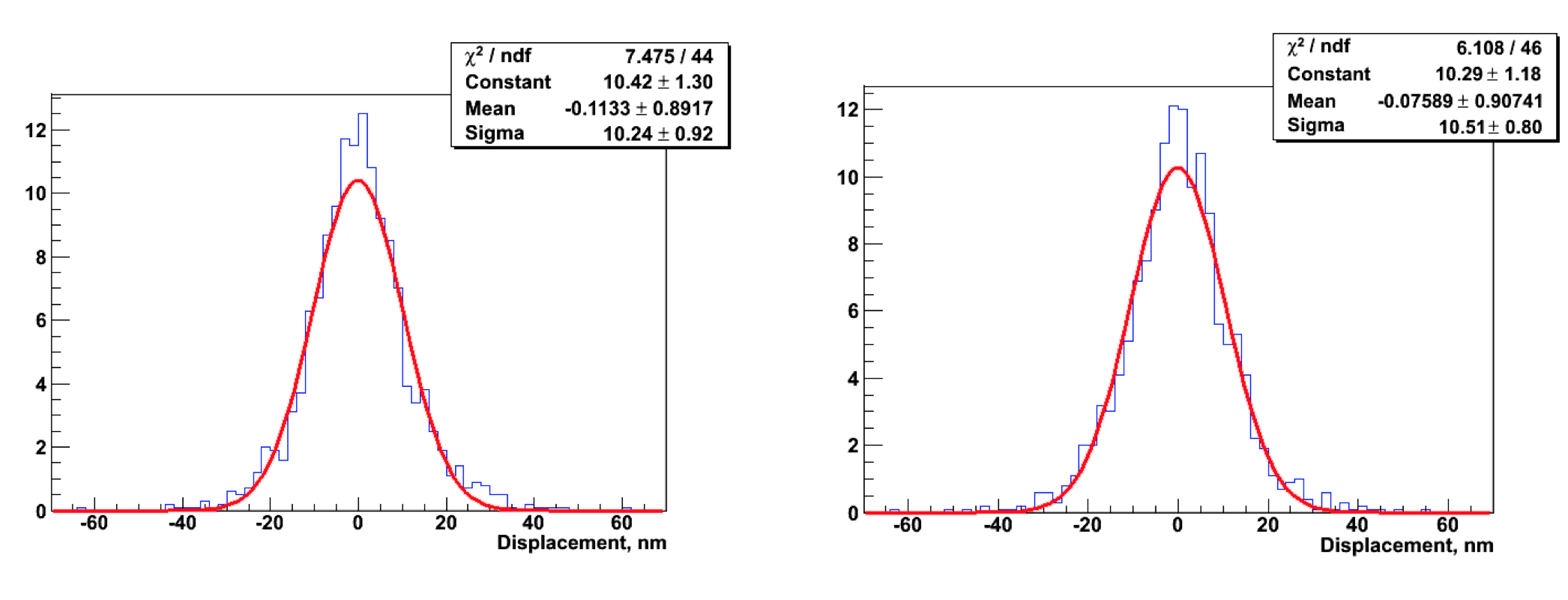}
	\caption{Position accuracy of $x$ (left) and $y$ (right) coordinates of about 10 nm with the resonant light scattering.}
	\label{fig:plasmon_resolution}
\end{figure}

In order to study the polarized light effect, several tests have been performed on NIT samples exposed to 100 keV C ions. Optical microscopes have been equipped with a polarization filter as shown in Figure~\ref{fig:plasmon_prototype}. 
The polarization direction can be changed by rotating the polariser. The rotation is at the moment done by hand while its automation is being designed. 
Images of the same clusters were taken by  rotating the polarizer of 180$^\circ$ with steps of 10$^\circ$.  An example of the analysis performed on a cluster with ellipticity 1.27 is reported in Figure~\ref{fig:plasmon_analysis1}. For all the images, the displacement ($dx$, $dy$) of the cluster barycenter in $x$ and $y$ coordinates is measured in terms of pixel units (1 pixel $=$ 55 nm). A displacement exceeding the position accuracy of a single grain is the evidence for a cluster made of two consecutive grains and therefore produced by a signal track. From the analysis of $dy$ versus $dx$ it is possible to retrieve the track length and slope. In this case, the measured track length is 1.5 pixel, corresponding to about 90 nm. 

The evaluation of the position accuracy was performed by analysing images of single grains. The unprecedented accuracy of about 10 nm can be achieved in both coordinates, as shown in Figure~\ref{fig:plasmon_resolution}.

The test performed demonstrates that this technology is very promising and that it can replace the X-ray microscope. The resonant light scattering has, in fact, the big advantage to achieve a nanometric resolution with optical microscopes. The validation of the candidates identified by the shape analysis will be performed in the same scanning laboratory, without moving the samples to a dedicated laboratory for the X-ray analysis. Moreover, optical microscopes are characterized by a much faster scanning speed with respect to X-ray microscopes, since they profit of all the R$\&$D performed in the last decades both for the  OPERA and the NEWS experiments.

\section{Expected Background} \label{sec:expected-bkg}
	
	The final sensitivity of low-energy rare event searches is strongly limited by the background induced by radioactivity. Two main categories have to be taken into account: the environmental or external background and the intrinsic one. The flux of the former can be significantly reduced by placing the detector underground, to absorb the cosmic radiation, and designing an appropriate shield against the natural radioactivity. The latter is an irreducible source of radiation: it is therefore crucial to control the radioactivity of the materials used for the construction of both the detector and of the shield as well as of the structure of the apparatus.
	
	Background sources for dark matter searches are $\alpha$ and $\beta$ particles, $\gamma$-rays and neutron induced recoils, while NIT are essentially not sensitive to minimum ionizing particles (MIP).
	
	The main sources of $\alpha$-particles are U and Th radioactive chains and Radon. The $\alpha$-particles produced in those processes have energies of the order of MeV and their range in emulsion is of the order of tens of microns, by far longer than WIMP-induced nuclear recoils. $\alpha$-particles can therefore be identified 
	and discarded in the emulsions by an upper cut on the track length.  Anyway Radon progeny $^{214}$Pb, $^{214}$Bi and $^{210}$Bi emit energetic $\beta$ and $\gamma$ radiation. To prevent Radon contamination, the detector has to be kept sealed from the air and continuously flushed with boil-off nitrogen.
	
\begin{figure}[tbph]
	\centering\includegraphics[width=0.7\linewidth]{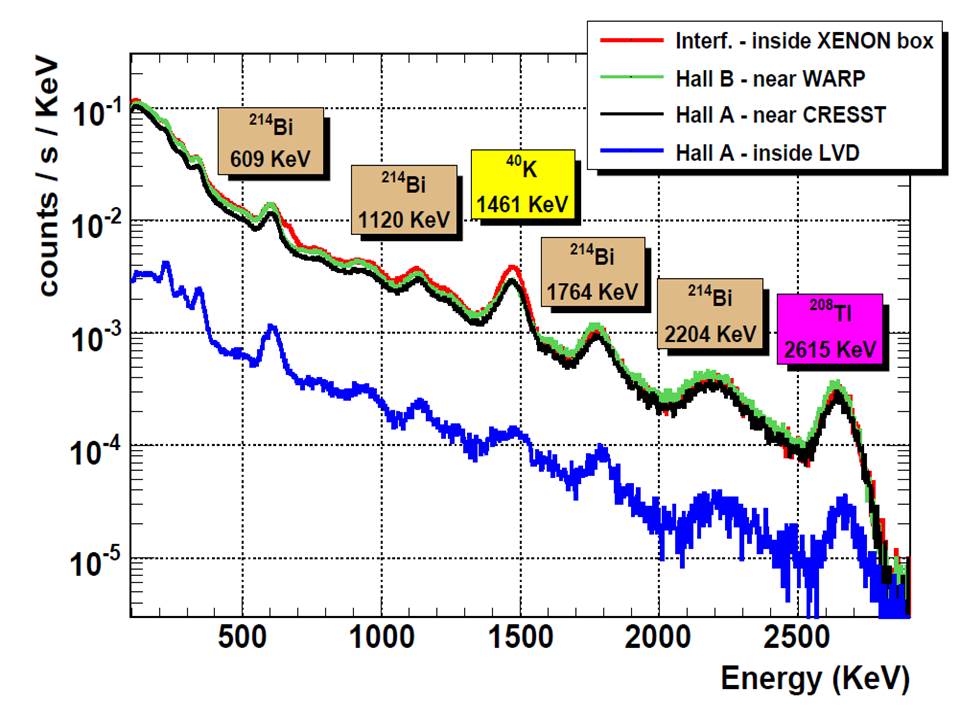}
	\caption{$\gamma$-flux measured in the underground LNGS halls \cite{BrunoPhDThesis,arneodo,wulandari}.} \label{fig:gamma-bkg}
\end{figure}

	The $\gamma$ radiation due to environmental radioactivity constitutes a non-negligible contribution to the total background budget. In Figure~\ref{fig:gamma-bkg} the measured $\gamma$ flux in the LNGS underground halls is shown~\cite{BrunoPhDThesis,arneodo,wulandari}.  Passive or active shielding (usually water, copper or lead) can be used to suppress the external $\gamma$-radiation down to the level of ppb or ppt. The thickness \emph{l} required to reduce the external flux by a factor $f > 1$ can be estimated assuming exponential damping $\emph{l} = \lambda (E_\gamma) \times \log f$, where $\lambda (E_\gamma)$ is the energy-dependent attenuation length and $E_\gamma$ is the $\gamma$-ray energy.

	A relevant source of background is represented by $\beta$-rays produced in $^{14}$C decay. Given the carbon content in the emulsions and the $^{14}C$ activity, a rejection power R$_{\beta}\leq10^{-8}$ is required in order to make it negligible (i.e. less than one background track/kg/year). The current rejection power for tracks made by two crystals is  R$_{\beta}=10^{-6}$. In order to further improve the rejection, three possibile improvements are under investigation. The first one is based on the different energy deposition per path length of WIMP induced recoils and electrons~\cite{gamma-response}: the response of emulsions can be tuned by dedicated chemical treatments (e.g.~Tetrazorium compound~\cite{tetraz}). The second possibility is to exploit the response of $\beta$-rays to the polarized light scattering: indeed grains induced by $\beta$-rays might be less sensitive to polarized light. Finally, a reduction of the background can be achieved by performing a cryogenic exposure and by exploiting the phonon effect. Preliminary tests at $\sim 100$ K show an upper limit of R$_{\beta}<10^{-7}$ for tracks made by two crystals. 

	%It is anyway worth noting that the background due to $\gamma$ radiation is less critical for NIT emulsion with respect to other sources. The energy deposition per unit path length of WIMP-induced recoils is in fact expected to be  one (light nuclei) or two (heavy nuclei) orders of magnitude larger than the one due to electrons~\cite{gamma-response}. 
		
	%$\gamma$-rays and $\beta$-particles can therefore be rejected by properly regulating the emulsion response, in terms of number of sensitized crystals per unit path length (i.e.~the sensitivity), through a chemical treatment of the emulsion itself. With this method a rejection power of the order of $10^{5}$ has been already achieved. 
		
Neutron induced recoils rank as the main background source because they are not distinguishable from the expected WIMP signal, except for the isotropic angular distribution and for the typical track length, largely exceeding the range expected for WIMP-induced recoils. Indeed, while neutron-induced proton recoils can be as long as few hundred microns, %(see left plot in Figure \ref{fig:proton_recoils})
the maximum length of a WIMP-induced nuclear recoil is smaller than $1\mu$m even for large ($O$(TeV)) WIMP masses. %(see Figure \ref{fig:maximumRange-vs-WIMPmass}). 
Three types of neutron sources affect underground experiments: radiogenic neutrons in the MeV range produced in ($\alpha$, n) and spontaneous fission reactions in the detector due to its intrinsic radioactive contaminants, cosmogenic neutrons with energy spectrum extending to GeV energies induced by muons penetrating underground through the rock, neutrons induced by environmental radioactivity.

\begin{figure}[tbph]
	\centering\includegraphics[width=0.7\linewidth]{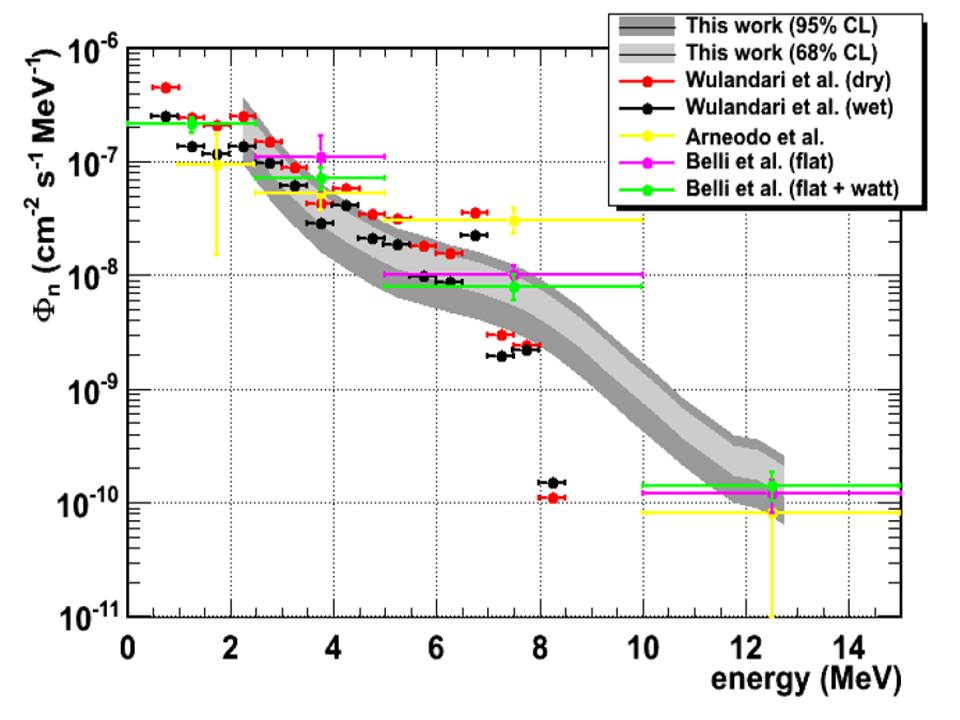}
	\caption{The neutron flux measured in the underground LNGS halls \cite{BrunoPhDThesis}.} 
		\label{fig:neutron-flux}
\end{figure}

In Figure~\ref{fig:neutron-flux} the measured neutron flux in the LNGS underground halls is shown \cite{BrunoPhDThesis}: for a neutron energy of the order of a few MeV (\emph{fast} neutrons) the flux ranges from $10^{-6}$ to $10^{-10}$ cm$^{-2}$ s$^{-1}$ MeV$^{-1}$. Light materials are effective moderators for fast neutrons: polyethylene  (PE, C$_2$H$_4$) is commonly used to reduce the external neutron flux. 

While the external neutron flux can be reduced to a reasonable level with an appropriate shielding, the intrinsic emulsion radioactivity would be responsible of an irreducible neutron yield through ($\alpha$, n) and $^{238}$ U spontaneous fission reaction. In order to estimate this contribution, a sample of each component of the nuclear emulsion (AgBr, Gelatin and PVA) has been analysed by the Chemistry Service in Laboratori Nazionali del Gran Sasso (LNGS, Italy), with the Inductively Coupled Plasma Mass Spectrometry (ICP-MS) technique~\cite{ICP-MS} and at the low background facility STELLA (SubTErranean Low Level
125 Assay) of the LNGS~\cite{STELLA} with germanium detectors. The complementary use of these techniques allows to determine both the Uranium and Thorium activities and to verify the secular equilibrium hypothesis. 

The measured activities are reported in Table~\ref{tab:activities-MS} for all the constituents. The upper limits on PVA are evaluated at 95$\%$ CL. 

By weighting the measured activity of each constituent for its mass fraction, the total activity of nuclear emulsion can be calculated. Using the contamination measured with the mass spectrometry, the $^{238}$U activity amounts to $23\pm 7$ mBq kg$^{-1}$, while the $^{232}$Th  one is $5.1\pm 1.5$ mBq kg$^{-1}$. The reported errors are dominated by the 30$\%$ uncertainty in the radioactive contamination measurements. By assuming a null contribution from PVA, the previous contaminations are reduced by $\sim 2\%$. 

The $\gamma$ spectrometry gives comparable results for the AgBr sample. For the gelatin the measurements provide comparable results for the $^{232}$Th chain, while the measured concentrations of $^{226}$Ra in the $^{238}$U chain is about 20 times smaller than the parent isotope, with a measured value of $2.4\pm 0.6$ mBq kg$^{-1}$. This measurement suggests a break in the secular equilibrium of the decay chain at this point. Therefore the secular equilibrium is assumed for the upper part of this chain, using the activity measured by mass spectrometry, while, for the lower part, nuclides are considered in equilibrium with  $^{226}$Ra and the activity measured with $\gamma$-spectroscopy is used. The nuclear emulsion activity for nuclides of the  $^{226}$Ra sub-chain is therefore $15\pm 5$ mBq kg$^{-1}$~\cite{intrisicBkgPaper}. 

\begin{table}[tph]
\begin{center}
\begin{tabular}{c|c|c}
\hline
Nuclide         & Contamination [10$^{-9}$ g g$^{-1}$] & Activity [mBq kg$^{-1}$] \\
\hline
\multicolumn{3}{c}{AgBr-I} \\
\hline
$^{232}$Th         & 1.0                       & 4.1                 \\
$^{238}$U        & 1.5                      & 18.5                  \\
\hline
\multicolumn{3}{c}{Gelatin} \\     
\hline
$^{232}$Th         & 2.7                       & 11.0                \\
$^{238}$U        & 3.9                      & 48.1                  \\
\hline
\multicolumn{3}{c}{PVA} \\
\hline
$^{232}$Th         & $< 0.5$             & $< 2.0$             \\
$^{238}$U        & $< 0.7$             & $< 8.6$                  \\
\hline
\end{tabular}
\end{center}
\caption{Results obtained by ICP-MS in terms of contamination and activity for the different constituents of the nuclear emulsion.
The estimated uncertainty is 30$\%$. The upper limits on PVA are evaluated at 95$\%$ CL.}
\label{tab:activities-MS}
\end{table}

\begin{figure}[tbph]
\centering\includegraphics[width=0.7\linewidth]{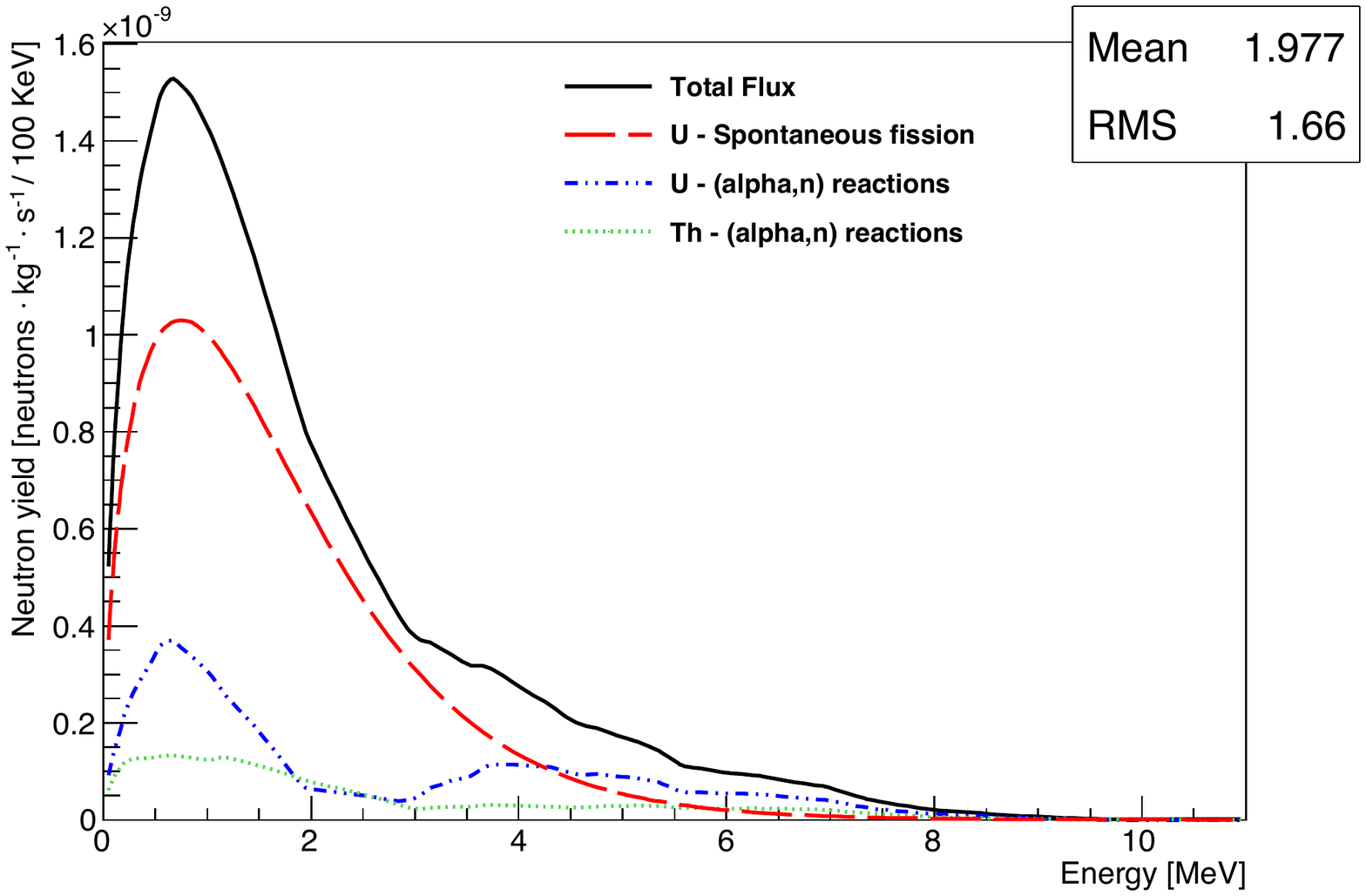}
	\caption{Total neutron energy spectrum (black line); in red the contribution from $^{238}$ U spontaneous fission is shown, while in blue and green the contributions from ($\alpha$,n) reactions due to nuclides in the $^{238}$U and $^{232}$Th chains respectively are displayed \cite{intrisicBkgPaper}.} \label{fig:SOURCES-spectrum}
\end{figure}

The measured activity was used to determine the neutron yield both through a semi-analitical calculation~\cite{refCalcFabio1,refCalcFabio2} and a MC simulation based on the SOURCES code~\cite{SOURCES}. Results are reported in Table~\ref{tab:resNeutronYield}. The two approaches give comparable results and the flux due to the intrinsic radioactive contamination is expected to be of the order of $1.2 \pm 0.4$ neutron per year per kilogram of nuclear emulsion. The energy spectrum of the produced neutrons, as calculated with SOURCES, is reported in Figure~\ref{fig:SOURCES-spectrum}.

\begin{figure}[tbph]
\centering\includegraphics[width=1.0\linewidth]{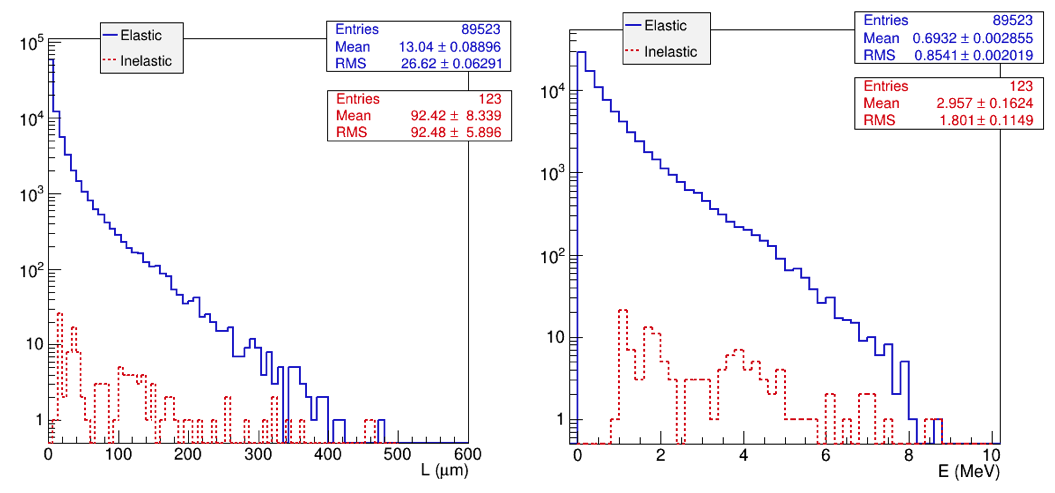}
	\caption{Track length (left) and energy spectrum (right) for proton recoils
produced by elastic (blue curve) and inelastic (red curve) processes.} \label{fig:proton_recoils}
\end{figure}

\begin{figure}[tbph]
\centering\includegraphics[width=1.0\linewidth]{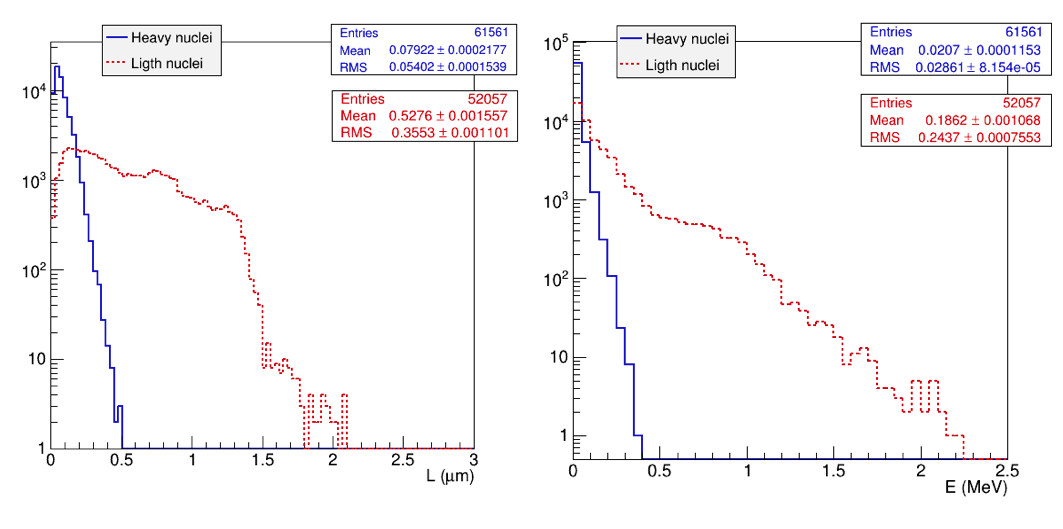}
	\caption{Track length (left) and energy spectrum (right) for heavy (blue curve)
and light (red curve) nuclei.} \label{fig:nuclear_recoils}
\end{figure}

\begin{figure}[tbph]
\centering\includegraphics[width=0.7\linewidth]{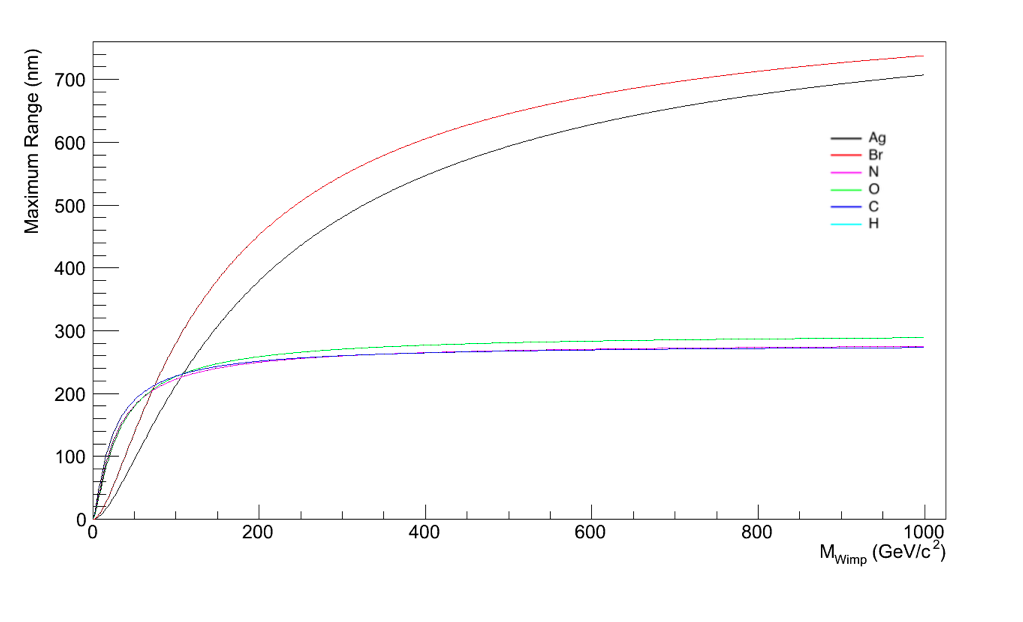}
	\caption{Maximum range expected for nuclear recoils as a function of the WIMP mass for the various nuclei.} \label{fig:maximumRange-vs-WIMPmass}
\end{figure}

In order to estimate the detectable background due to radiogenic neutrons produced by the intrinsic radioactive contamination of the nuclear emulsions, a GEANT4 based simulation was performed. Simulated neutrons have an isotropic angular distribution and are uniformly distributed in a target where emulsion are arranged in a stack with a surface of $25 \times 25$ cm$^2$ and a thickness of 0.5 cm; their energy spectrum was generated according to Figure~\ref{fig:SOURCES-spectrum}. The fraction of interacting neutrons is 20.4$\%$: they can produce either a proton a nuclear recoil. In the former case the track length in emulsion extend up to several hundreds $\mu$m (see Figure \ref{fig:proton_recoils}) while nuclear recoils show shorter track lengths, not exceeding 3 $\mu$m for light nuclei (C, N, O) and 1 $\mu$m for heavy nuclei (Ag, Br, I) (see Figure \ref{fig:nuclear_recoils}). The overall fraction of neutron-induced recoils contributing to the background is computed by accounting for recoil tracks with lengths above the read-out threshold. Moreover an upper limit on the track length can be introduced since the signal is expected to be below 1 $\mu$m even for large ($O$(TeV)) WIMP masses (see Figure \ref{fig:maximumRange-vs-WIMPmass}). The fractions of neutron-induced recoils below this cut, as a function of the read-out threshold, are reported in Table~\ref{tab:recoils1}: only fraction, from 5\% to 10\%, contributes to the background. A further reduction of $\sim 70\%$ of the neutron-induced background can be achieved exploiting the directionality information with the cut $-1 < \phi < 1$. Under these assumptions, the detectable neutron-induced background would be 0.02 $\div$ 0.03 per year per kilogram.

%\ref{fig:neutronRecoil_range})
%\ref{fig:maximumRange-vs-WIMPmass}

\begin{table}[tph]
\begin{center}
\begin{tabular}{c|c|c}
\hline
Process                               & SOURCES simulation         & Semi-analytical
calculation        \\
                                      & [kg$^{-1}$ y$^{-1}$]      & [kg$^{-1}$ y$^{-1}$] \\
\hline
($\alpha$, n) from $^{232}$Th chain   & 0.12                    & $0.10 \pm0.03$                \\
($\alpha$, n) from $^{238}$U chain    & 0.27                    & $0.26 \pm 0.08$                \\
Spontaneous fission                   & 0.79                    & $0.8 \pm 0.3$                \\
\hline
Total flux                             & 1.18                    & $1.2 \pm 0.4$                \\
\hline
\end{tabular}
\end{center}
\caption{Neutrons per kilogram per year due to ($\alpha$, n) and spontaneous fission
reactions in the
  nuclear emulsion, evaluated with the SOURCES code and semi-analytical calculation
using the measured $^{238}$U and $^{232}$Th contaminations as input.}
\label{tab:resNeutronYield}
\end{table}

The neutron-induced background due to the intrinsic radioactive contamination allows the design of an emulsion detector with an exposure of about 10 kg year. A careful selection of the emulsion components and a better control of their production could further increase the radiopurity, thus extending the detector mass and exposure time. In particular, since the activity of the gelatin is higher than that of the other emulsion components (see Table~\ref{tab:activities-MS}) and since PVA shows a very low radioactive level, we are studying a possible replacement of  gelatin with PVA.

In nuclear emulsion-based detectors the instrumental background is due to the so called \emph{fog} grains, i.e.~dark grains produced by thermal excitation. The fog density determines the probability of random coincidences of two or more fog grains mimicking a WIMP-induced nuclear recoil. The measured value of the fog density for current NIT samples is about 0.1 grains/(10$\mu$m)$^3$.

\begin{figure}[tbph]
	\centering\includegraphics[width=1.0\linewidth]{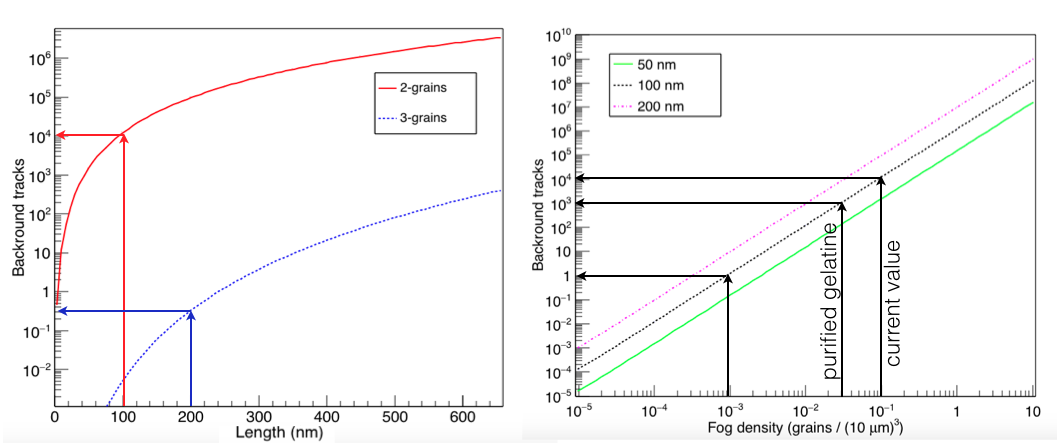}
	\caption{Left: number of background tracks in 1 kg of NIT emulsions as function of the track length for tracks made by two  (continuous red  line) and three fog grains (dashed blue line). Right: number of background tracks in 1 kg of NIT emulsions as function of the fog density for 50 nm (continuous green line), 100 nm (dashed black line) and 200 nm (dotted-dashed magenta line) threshold in the track length. Only tracks made by two grains are considered here. }
	\label{fig:combinatorial_bkg}
\end{figure}

The number of background tracks due to random coincidences of fog grains depends on the minimum number of grains required to build a track and increases with the track length, as shown in the left plot of Figure~\ref{fig:combinatorial_bkg}, where the instrumental background for 1 kg emulsion target is reported. 
In NIT (U-NIT) emulsions a track made of 2 grains has an average length of about 100 nm (50 nm). The number of background tracks corresponding to this track length amounts to 10$^4$ (10$^3$), as outlined by red arrows on the plot.  
\\
In order to make the combinatorial background smaller than one, the coincidence of at least 3 grains has to be required. In NIT (U-NIT) emulsions a track made of a sequence of 3 grains has on average a path length of about 200 nm (100 nm): the corresponding background level is 0.3 tracks ($4\times10^{-3}$ tracks).
\\
The right plot in Figure~\ref{fig:combinatorial_bkg} shows the number of background tracks as function of the fog density in NIT emulsions, if 2-grain tracks are accepted: the background can be considered as negligible only reducing the fog density to from the current value to 10$^{-3}$~grains/(10$\mu$m)$^3$. Preliminary tests show that a value of 0.03 grains/(10$\mu$m)$^3$ can be obtained using purified gelatine. Further purification might lead to lower fog values. This research line will be followed in collaboration with the firm producing the gelatine.
\\
In order to further reduce the fog density, two possible improvements are under study.
The first one exploits the response of fog grains to the polarized light scanning:
fog grains show indeed both a different image contrast and a different size with respect to the grains sensitized by a nuclear recoil. This effect is essentially due to different $dE/dx$ of the two processes and offers a powerful discrimination of such kind of background.
Moreover, a reduction of the fog density can be achieved operating the detector at low temperature (from a simple refrigeration down to a cryogenic regime of $\sim 80$~K) or by applying dedicated chemical treatments. 

%\newpage

Finally, the requirement of a background-free experiment sets the necessity of operating in a clean environment in order to avoid surface contamination. 
Moreover in order to reduce the activation risk of detector materials, an underground location for the emulsion production and handling facilities is required. The construction of a (dark) clean room in the Gran Sasso underground Laboratory is therefore needed. 
	
\begin{table}[tph]
\begin{center}
\begin{tabular}{c | c }
\hline
  Threshold [nm] &  Fraction \\
  \hline
 50  & 0.100   \\
 100 & 0.075   \\
 150 & 0.060   \\
 200 & 0.052   \\
 \hline
 \end{tabular}
 \end{center}
 \caption{Fraction of detectable neutron-induced recoils as a function of the
read-out threshold.}
 \label{tab:recoils1}
 \end{table}	

\section{Experimental set-up} \label{sec:set-up}

As a first phase of the project, we plan to perform a pilot experiment with a detector of 1 kg exposed for one year. Details of the related schedule will be examined in Section~\ref{sec:schedule}. 

A detector with one kg mass of NIT can be made of 50 $\mu$m thick-films assembled in a stack of 100 planes with a surface of $25 \times 25$ cm$^2$. %Each plane is made of 25 NIT emulsion films with dimension 5$\times$5 cm$^2$. 
We are considering the option of embedding  OPERA-like films between two consecutive NIT planes: OPERA-like films would act as a monitoring system to register, with micrometric accuracy and high sensitivity, all the radiation integrated by the detector along the exposure. As composed of the same raw materials, the intrinsic radioactivity of the OPERA-like films would be of the same order of magnitude of that of NIT, therefore tolerable for a 1 kg detector. 
	
	The emulsion planes are placed with their surface parallel to the expected WIMP wind direction. We might consider to place an equivalent amount of emulsion films in an orthogonal plane. These films would act as a control sample. In case a signal would be found in thefirst sample, and only in this case, the scanning of these films would be performed to demonstrate that the signal found is not an artefact.

\begin{figure}[tbph]
	\centering\includegraphics[width=0.7\linewidth]{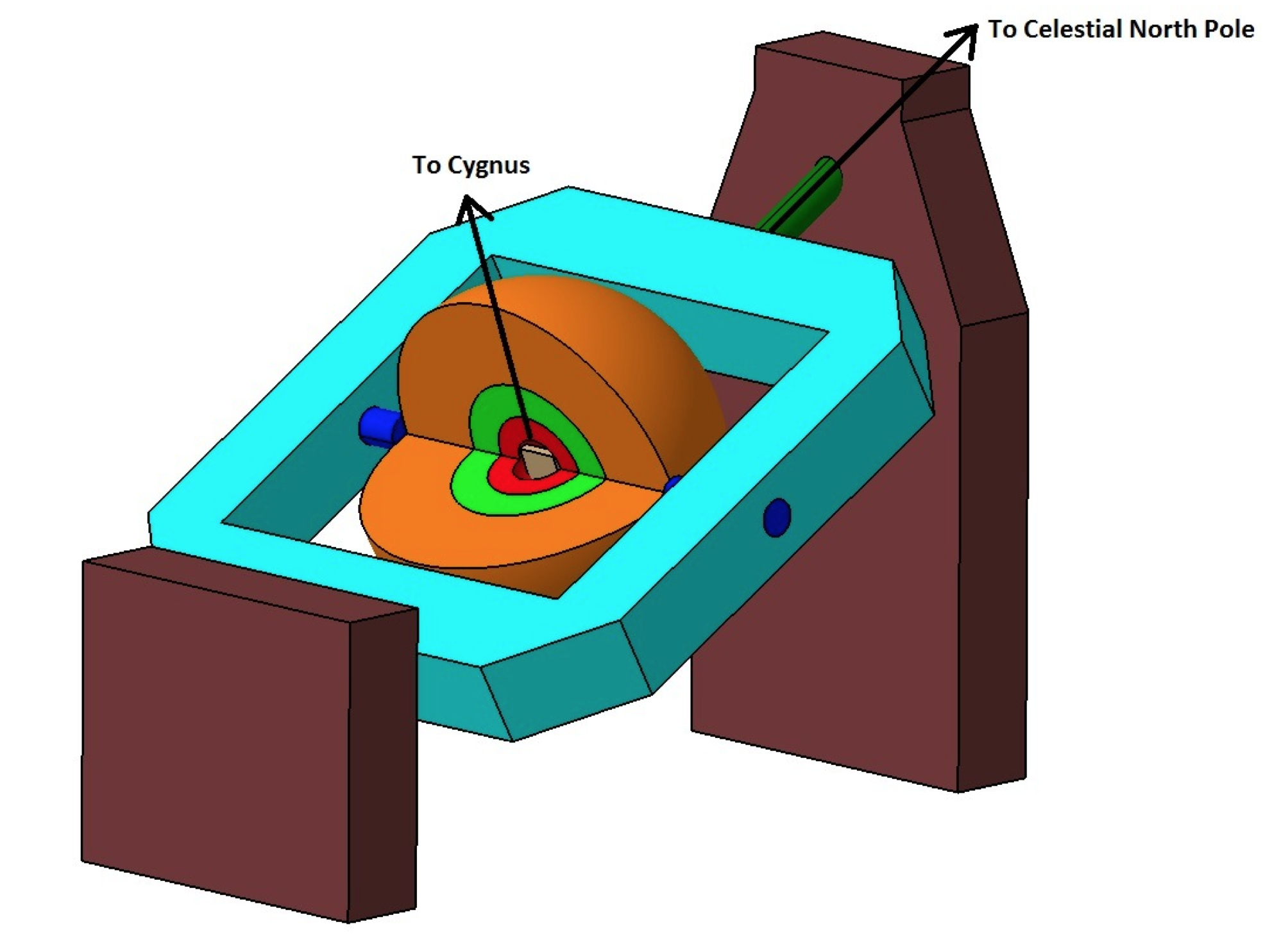}
	\caption{Schematic view of the detector structure.} \label{fig:detector}
\end{figure}

	To maintain the detector with a fixed orientation towards the Cygnus constellation it will be installed on an Equatorial Telescope (see Figure~\ref{fig:detector}) allowing to cancel out the effect of the Earth rotation  thus keeping the detector pointed on a fixed position in the sky. 
	
An equatorial telescope has two axes: the so called Polar Axis, parallel to the rotation axis of the Earth and pointed to the North celestial pole, and the Declination Axis, perpendicular to the polar one. The motion of the Earth can be canceled out by driving at a constant speed the Polar Axis synchronised with the apparent daily motion of the sky. The Polar Axis will be motorized and both axes will be equipped with precise encoders to constantly check the position of the mechanics with high accuracy. The detector will be therefore pointed towards the Cygnus constellation and kept in that direction with an accuracy better than 1 degree.

A calibration procedure of the telescope will be performed before the installation in the underground laboratory. To ensure a precise syncronization of the mount with the apparent daily motion of the sky it is necessary to tune the response of the mechanics and to correct for any possible periodic error. The calibration procedure foresees several steps. The mount will at first be tested in the external laboratory using an optical telescope mounted on it and aligned with the Polar Axis: the telescope will be used, during the night, to point a star in the Cygnus constellation. Using a specific software and an imaging CCD camera attached to the prime focus of the telescope, the mount will be guided to keep the star centered in the field of view of the CCD camera. The software will record all the guiding parameters, as the position of the star and the corrections applied to the Polar and Declination Axis. This procedure will be repeated during several nights and all the data collected will be analyzed in order to get and apply the necessary correction to the mechanics and to the electronic system.

In a second phase the mount will be used throughout the whole day to compensate the apparent daily motion: the position during the night will be then measured in order to evaluate the pointing accuracy given by the 
difference between the nominal and measured positions. This measurement will provide a fine tuning on the position of both the Polar and the Declination axes.  

Finally the mount will be moved underground in its final position: profiting of the presence, in the underground halls, of already existing high precision reference points the mount will be aligned with high accuracy in the north-south direction in order to align the Polar Axis parallel to the rotation axis of the Earth.

The design and construction of the equatorial telescope  will be carried-out in collaboration with specialized firms. A screening of all the materials used in the construction of the telescope is foreseen in order to evaluate their intrinsic radioactivity.  A detailed simulation of all the components of the telescope is planned. 

A large telescope supporting both the target and the surrounding shield is considered  (see Figure~\ref{fig:detector}). This configuration allows to build a light shield while ensuring a low contamination of the background originating from the telescope itself. 
	
	In Figure~\ref{fig:detector} a schematic view of the detector structure is shown: a stack of NIT films is placed at the center of a plexiglass sphere with a diameter of 30 cm.  A sphere of polyethylene will act as a shield against the external neutron background. The target and the shielding are installed on the equatorial telescope. The target emulsions are arranged in such a way to have  the film surface parallel to the WIMP wind. 

From a preliminary evaluation a layer of 50 cm of polyethylene will reduce the external neutron flux by a factor of the order of $10^{4}$: considering an integrated flux of the order of $\phi_n \sim 2 \times 10^{-6}$ cm$^{-2}$ s$^{-1}$, for a target with an exposed surface of $25 \times 25$ cm$^2$ and a thickness of 0.5 cm this corresponds to a residual flux of the order of 1 neutron/kg/year, the same order of magnitude of the intrinsic neutron contamination. 
More accurate evaluations of the polyethylene thickness sufficient to provide the required background rejection power are under study.
The addition of a thin ($1\div 2$ cm) layer of Cadmium to capture thermalised neutrons is under study. 

As explained in Section~\ref{sec:expected-bkg} NIT have a high electron rejection power: a proper chemical treatment allows to reach a reduction factor of the order of 10$^{-6}$ in the sensitivity to electrons. For this reason the use of high-Z shielding materials (Pb and Cu) against the external $\gamma$ flux is not foreseen at the moment.

Both the passive shield and the emulsion target will be enclosed in a sealed plexiglass box maintained in High Purity (HP) Nitrogen atmosphere in slight overpressure with respect to the external environment to prevent radon contamination.

The shape of the shield surrounding the detector will be optimized in order to design the lighter and efficient structure. Two solutions are under study, either a parallelepiped box containing the shielding and the emulsion target, or a spherical one. In the first case the weight of the PE layer is of the order of 1.8 ton, while in the spherical option the weight is $\sim$1.14 ton. Even if the latter option ensures 
a lighter and symmetrical  shielding, the final choice will depend on the cost and on the technical implementation of the design. 

\begin{figure}[thp]
	\centering\includegraphics[width=1.0\linewidth]{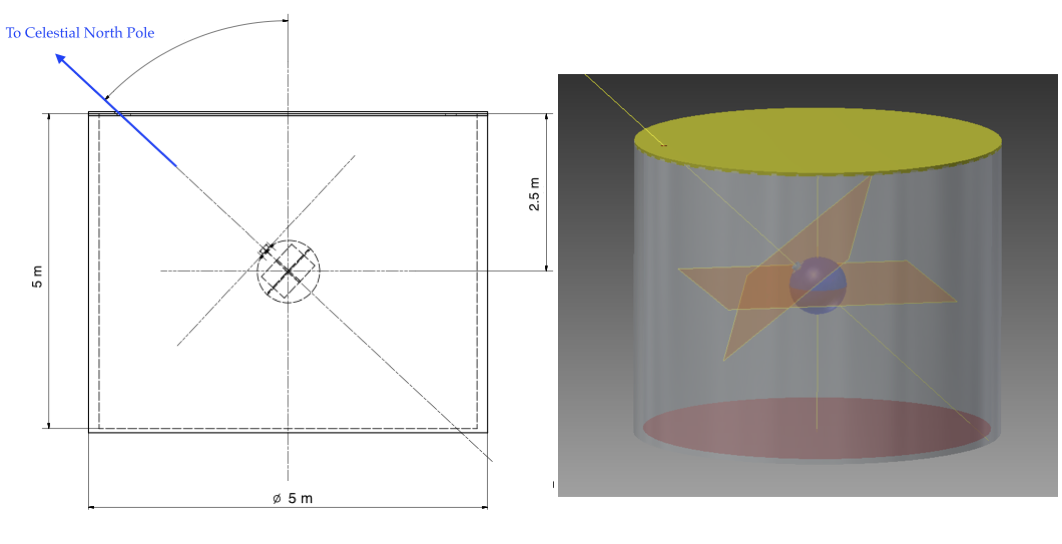}
	\caption{Schematic view of the detector structure for the water shielding: the detector holder is placed in the centre of tank. Its orientation toward the Cygnus constellation is kept by the rotating pivot mounted with one edge above the water surface. Only pure and low-Z materials are used for the immersed part.} \label{fig:WaterOption}

\end{figure}

Nevertheless, we are also investigating a completely different approach based on the use of water as shielding material against the external background. A preliminary layout is shown in Figure~\ref{fig:WaterOption}: the emulsion detector is hermetically enclosed inside a spherical container made of low-Z material (teflon or polyethylene) with a diameter of 55~cm. The inner volume is flushed with N$_2$. The container is mounted on a long shaft  and positioned in the centre of  a tank (diameter 5~m, height 5~m) filled with ultra-pure water. The shaft is made of light,  low-radioactive material (i.e.~aluminum)  aligned with Earth's rotation axis.  The constant orientation of the target with respect to the Cygnus constellation is kept thanks to the slow rotation of the shaft with the period of one sidereal day. All  the mechanics needed to keep the orientation and the rotation of the shaft is mounted outside the water tank.  The immersed part can be constructed in a way to keep the mean density close to 1~g/cm$^3$.  In this way the mechanical load becomes negligible, thus simplifying the design and providing a big flexibility for materials selection.

This solution can be more flexible and cheaper, allowing to hold much larger masses without changing neither the mechanics of the telescope nor the shielding. A detailed simulation of the shielding and a study of the mechanics requirements, together with an estimation of the costs, are ongoing.

\begin{figure}[thp]
	\centering\includegraphics[width=0.85\linewidth]{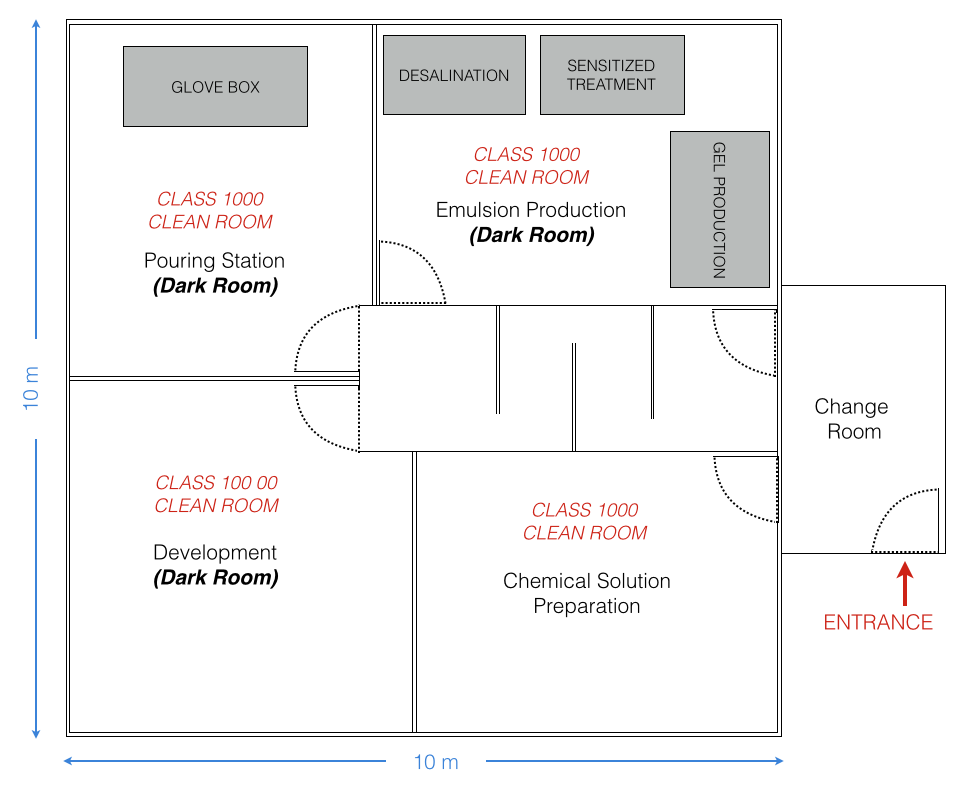}
	\caption{Sketch of the planimetry of the NIT production and development facility.} \label{fig:emulsion-facility}
\end{figure}

\begin{figure}[thp]
	\centering\includegraphics[width=0.7\linewidth]{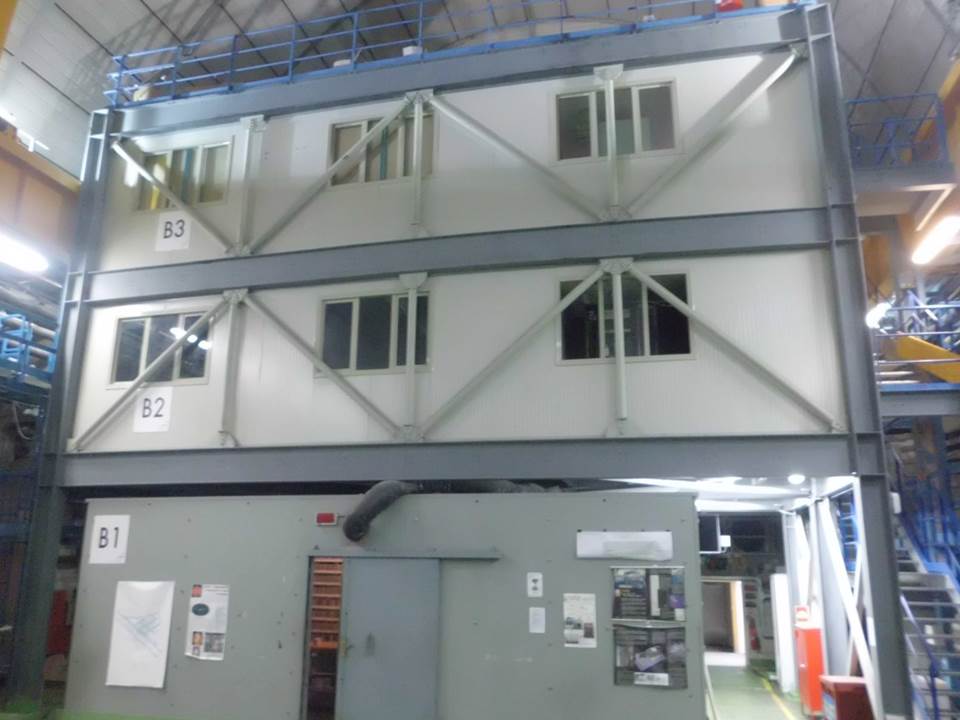}
	\caption{A picture of the existing OPERA CS facility in hall B.} \label{fig:CSemulsion-facility}
\end{figure}

\begin{figure}[thp]
	\centering\includegraphics[width=0.7\linewidth]{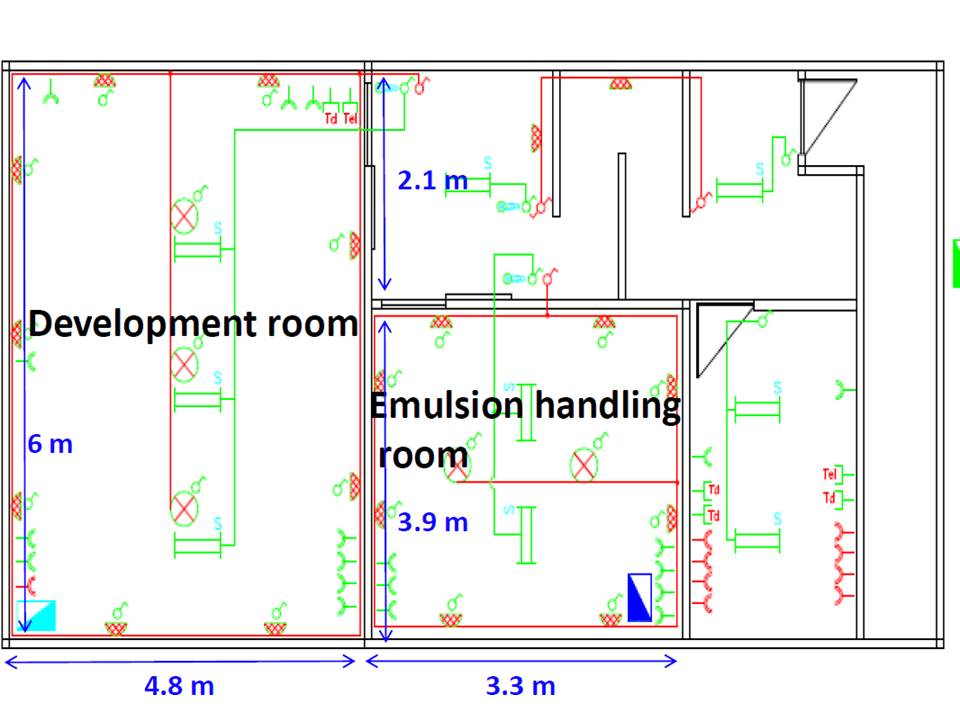}
	\caption{Planimetry the existing OPERA CS facility in hall B.} \label{fig:CSemulsion-facility-planimetry}
\end{figure}

\subsection{Emulsion production and development facility}

The layout of the facility we intend to build is shown in Figure~\ref{fig:emulsion-facility}. The total surface is about 100 m$^2$ and it is divided in four parts: emulsion gel production, emulsion gel pouring, film development and chemical solution preparation.\\
Once produced, the gel will be sealed in an envelop flushed and filled with N$_2$ and moved to the pouring station, where  a glove box flushed with HP Nitrogen will be installed. After the pouring, the films will be sealed in an envelop flushed and filled with N$_2$ and stored underground until the exposure. \\
All the operations involving the emulsion production and development require a dark room environment.\\
In order to minimize the surface contamination and the activation risk, the facility will be hosted in a clean room located underground. A class 1'000 clean room is required for the emulsion production, pouring and the chemical solutions preparation; a class 100'000 will be installed for the area devoted to the development.\\
An air conditioning system will be installed in order to stabilize and monitor the temperature, ($20 \pm 1)^\circ$, and the humidity, ($60\pm 5)\%$, of the clean room. A demineralized water treatment plant and a chemical waste system are also required. 

For the film development and  the pouring activity foreseen in the first year of the project an 
 excellent starting point is the existing OPERA emulsion handling facility shown in Figure~\ref{fig:CSemulsion-facility}. The facility, currently hosted in Hall B, is made of three rooms: a control room, a handling room and a development room, for a total surface of $\sim$ 50 m$^2$ (see Figure~\ref{fig:CSemulsion-facility-planimetry}). The handling room will be equipped with a pouring station and a  development station.
The installation of two systems for the temperature control is also foreseen.

The scanning of the exposed films will be performed in the existing OPERA scanning facilities in Italy, Russia, Turkey and Japan. In Italy the scanning laboratories are located at LNGS, Naples and Bari with 13, 5 and 3 OPERA microscopes respectively. Few more microscopes are currently located in Russian and Turkish scanning laboratories. Moreover at LNGS and Naples two additional microscopes, partially upgraded for the scanning of NIT and the polarized light analysis, are available. An equivalent scanning power is hosted at Nagoya University. 

\section{Physics reach}
	
\begin{figure}[hbtp]
	\centering
		\includegraphics[width=0.6\linewidth]{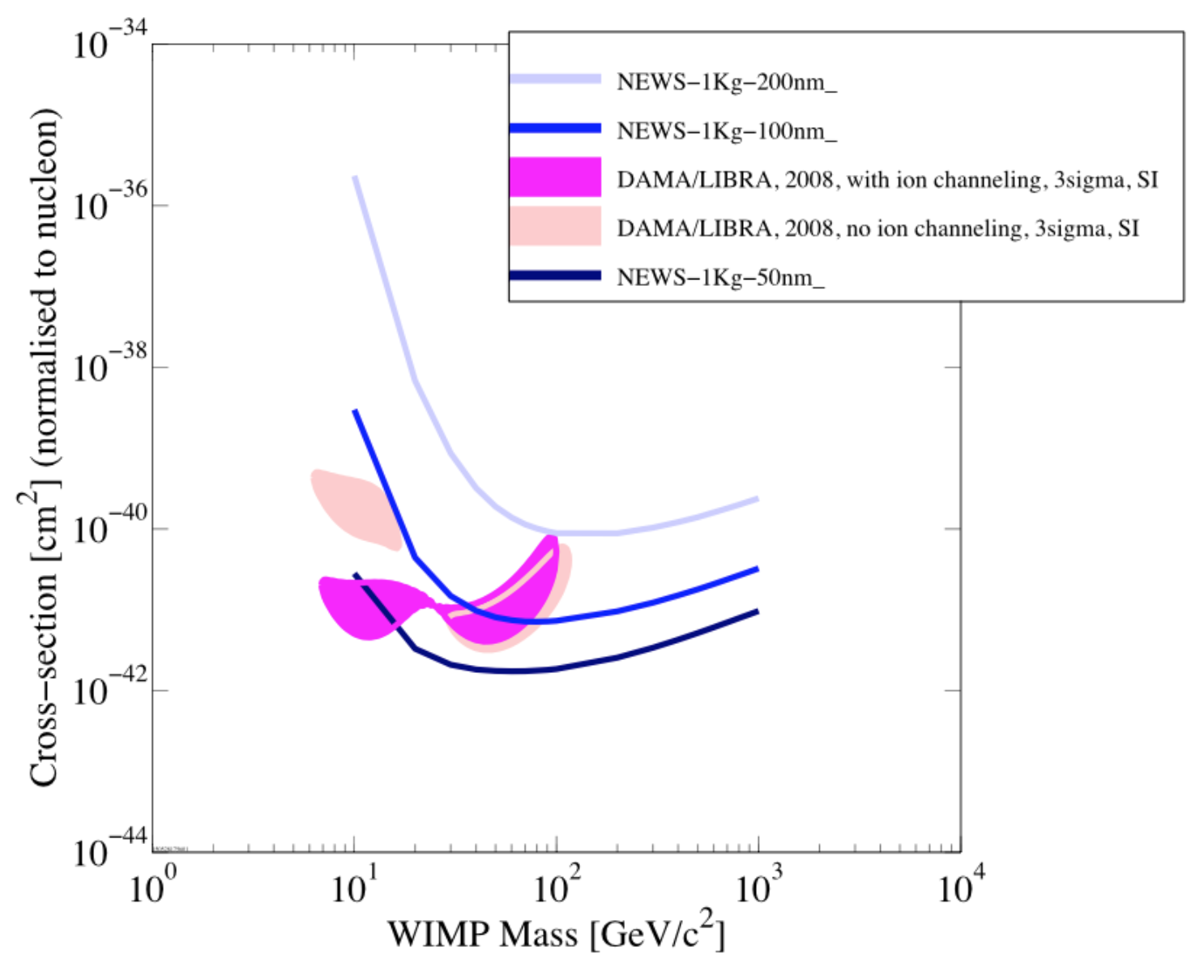}
	\caption{The 90$\%$ C.L. upper limits for a NIT detector with an exposure of 1 kg $\times$ year, a threshold ranging from 200 nm down to 50 nm, in the zero background hypothesis. The directionality information is not included.}
	\label{fig:sensitivity1Kg}
\end{figure}

	The 90$\%$ C.L. upper limit in case of null observation is shown in Figure~\ref{fig:sensitivity1Kg}
	for an exposure of 1 kg$\cdot$year of NIT emulsions, with a minimum detectable track length ranging from 200 nm down to 50 nm and in the hypothesis of zero background. Even not including the directionality discrimination of the signal and assuming to reach a negligible background level, such an experiment would cover a large part of the parameter space indicated by the DAMA/LIBRA results with a small (1 kg) detector mass, using a powerful and complementary approach. 
	
	It is worth noting that the sensitivity strongly depends on the final detection threshold: as explained in Section~\ref{sec:expected-bkg} the current threshold value is limited to 200 nm only by the fog density. A reduction of the fog density or its discrimination through the use of the optical microscope with polarized light, would allow to lower the threshold to 100 nm. In order to lower the threshold down to 50 nm the use of the U-NIT technology is needed. Moreover we are conservatively assuming zero efficiency below the threshold value while, as shown in Figure~\ref{fig:eff_vs_length}, the efficiency is not negligible even for shorter tracks. This would enhance the sensitivity to low WIMP masses. This effect will be taken into account.

\section{Schedule, Cost Estimate, Organization}  
	
	\subsection{Time schedule} \label{sec:schedule} 
	
	\begin{sidewaysfigure}[hbtp]
    \includegraphics[width=1.2\linewidth]{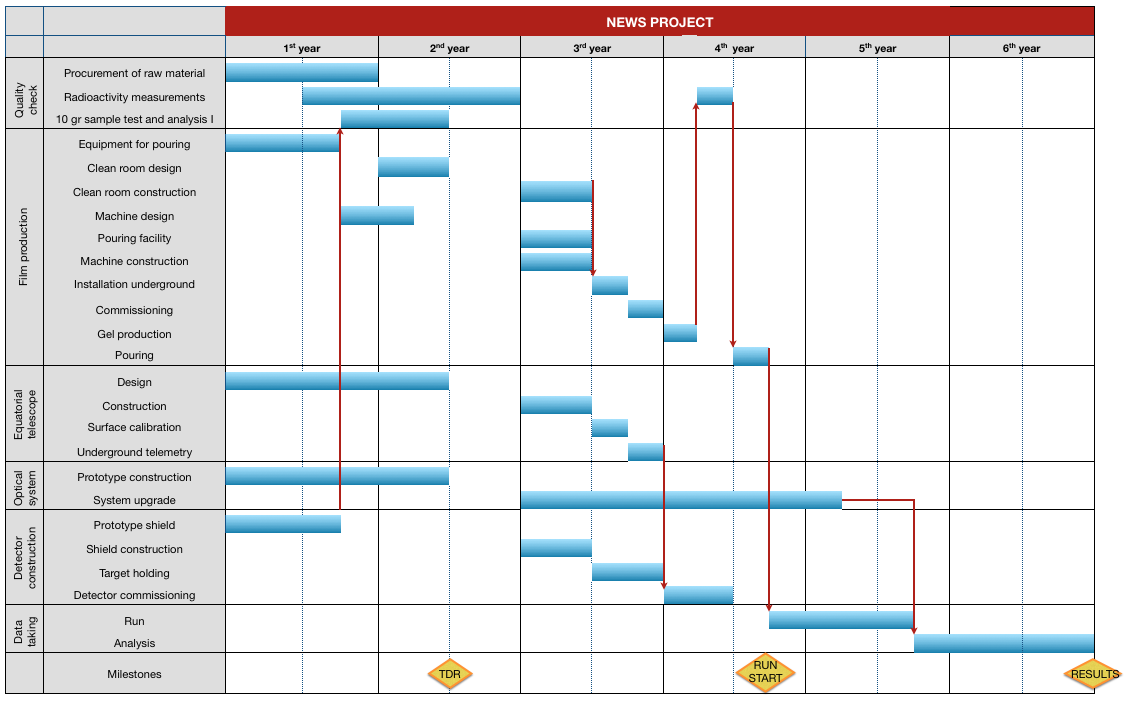}
    \caption{Gantt diagram with the different phases of the project.} \label{fig:gantt}
\end{sidewaysfigure}

	On a time scale of six years we intend to perform the first exposure with a target mass of 1 kg and the corresponding analysis of the data taken. In Figure~\ref{fig:gantt} a detailed plan of all the phases of the project is reported.  

	In the beginning of 2016 we plan to construct a prototype shield and the equipment for the emulsion pouring. The above mentioned phases  have to be completed in nine months in order to perform a first test to benchmark the level of intrinsic radioactivity of emulsions. For this measurement, we will use the gelatine produced at Nagoya University and perform the pouring underground. We will perform an exposure of a 10 g detector surrounded by the prototype shield. The detector exposure together with the analysis of the emulsion films will last nine months.
The results of this test will provide a measurement of the background, intended to cross-check the estimates based on simulation and measurements of intrinsic radioactivity. In parallel, tests with radioactive sources are foreseen to characterize the response to external radioactivity. 

	We do consider the possibility of getting raw materials for the emulsion production within European countries,
	provided that their intrinsic radioactivity does not exceed the level measured in Japanese samples. 
	This would allow a reduction of the activation processes induced during transportation. 
	This activity will take place in 2016.
	
	The measurement of  intrinsic radioactivity of the different emulsion components and the prototype shield materials will be performed from June 2016 to the end of 2017.
	Tests of the activation due to cosmic rays during the transportation will be performed by bringing samples back and forth between Italy and Japan. 

The design of the gel production machine will  start in September 2016 while the design of the clean room will be carried on with the help of specialized firms, starting from January 2017.

	%The phases related to the gel production machine will take place only if measurements of the gel radioactivity will show that the one produced in the Gran Sasso underground laboratory has a radioactivity level smaller than the one produced in Japan.  

The construction of the clean room, the pouring facility and the gel production machine will start from January 2018 and will last six months.
		As soon as the film production machine will be operational in the underground laboratory and the gelatine will be produced, the measurement of  intrinsic radioactivity will be performed. If satisfying the required radioactivity level, the pouring of the gelatine will be performed.
	
	%The phase related to the installation of the gel production machine in the LNGS underground facility will take place only if the tests of the activation risk due to cosmic rays during the transportation, will demonstrate an encrease in the radioactive level.  Such tests will be performed in 2016 by bringing gel samples, produced at Nagoya University,  back and forth between Italy and Japan. 
	%On the contrary, if such tests will show that the activation risk is negligible we will consider the possibility to do not move the gel production from the Nagoya University thus reducing significantly the space needed for the clean room facility in the LNGS underground hall. Indeed, in this hypothesys, only the pouring facility and the development room will be realized. 
	
	The design of the equatorial telescope and the choice of the materials is supposed to start soon in 2016 and last 18 months. The construction of the telescope will start in the beginning of 2018 and last six months. In the second part of  2018 the surface calibration measurements and the underground telemetry will be carried out. 

	The construction of the shield and the target holding will start in 2018. Once the equatorial telescope installation will be finalised, the detector commissioning will start.

	We plan to finalize the upgrade of the read-out system on a  prototype microscope, exploiting in particular the resonant light scattering technique. This activity will start at the beginning of 2016.
	
In June 2017 a Technical Design Report will be submitted.
	
The upgrade of all the available OPERA systems will start in the second half of 2018 and last 27 months.
By  March 2020 we plan to have the final equipment installed on a number of microscopes adequate for the analysis of 1 kg of NIT emulsion in one year. %The required number of fully automated microscopes is estimated to be ranging from 10 to 14 depending on the final achieved scanning speed at nanometric resolution.  

	Once the whole film production will be completed, the run with 1 kg mass detector will start. The data taking will last one year: from October 2019 to October 2020.  The emulsion films will be developed soon after the exposure. 
The scanning and the analysis of the emulsion films will start once the upgrade of all the read-out systems will be complered and it is supposed to be completed by the end of 2021.

	\subsection{Costs}

	The cost for the constructions of the clean room (75 m$^2$ class 1'000 and 25 m$^2$ class 100'000)  is estimated to be around 200 k\officialeuro. As explained in Section~\ref{sec:set-up} the clean room will host the production machines, the pouring and the development facilities.\\
The cost of the production machine is of the order of 200 k\officialeuro.
The pouring and the development facilities will cost about 18 k\officialeuro and 50 k\officialeuro, respectively.
The above mentioned costs  will be shared according to a MoU to be signed between parties. In case the production will be carried out at Nagoya University, Japan will cover the corresponding costs. Japan will cover the costs for the all the emulsion components. \\
A first estimate of the cost for the  equatorial telescope is 15 k\officialeuro
 for the design and 240 k\officialeuro; the cost for  the construction of the shielding amounts to about 15 k\officialeuro. \\
Finally the upgrade of the read-out systems will be needed. Japan will cover the cost for the realization of their own scanning systems.
The construction of the microscope prototype in Europe costs about 300 k\officialeuro;
the hardware and computing upgrade of each OPERA microscope amounts to about 30 k\officialeuro. Depending on the final scanning speed, from 10 to 14 systems will be modified for the high resolution and high speed scanning of NIT for a total cost ranging from $\sim$ 300 k\officialeuro to $\sim$ 420 k\officialeuro. \\
An expense of 80 k\officialeuro is expected for the maintenance of the microscopes and 120 k\officialeuro for the consumables.

In Table \ref{tab:costs} a summary of the expected costs is reported.

\begin{table}[htb]
\centering
\begin{tabular} {c | c | c} 
\hline
\hline
Category  & Cost  [k\officialeuro] &  	Assignment	\\
\hline
Clean Room  &  200  &  EU \\
NIT production machine   &  200 &  JP  \\
Pouring facility   &  18 &  EU  \\
Development facility   &  50 &  EU  \\
Equatorial Telescope & 255  & EU \\
Shielding &  15 &  EU\\
EU Prototype Microscope  & 30 &   EU    \\
EU Microscopes Upgrade  & 300	&   EU    \\
EU Microscope Maintenance  & 80 &   EU    \\
JP Microscopes Upgrade  & 300	&   JP    \\
Consumables & 120 & EU\\
\hline
TOTAL  & 1468 & \\
\hline
\hline
\end{tabular}
\caption{Summary of the expected costs.}
\label{tab:costs}
\end{table}

	\subsection{Collaboration}
	
	NEWS is  at present a collaboration between Italy, Japan and Russia and Turkey.
	
	The involved groups are:
	
	\begin{itemize}
		\item University and INFN Bari, Italy
		\item Lab. Naz. Gran Sasso, Italy
		\item University and INFN Naples, Italy
		\item University and INFN Rome, Italy
		\item Nagoya University and KM Institute, Japan
		\item Chiba University, Japan
		\item JINR Dubna, Russia
		\item Moscow State University, Moscow, Russia
		\item Lebedev Physical Institute, Moscow, Russia
		\item METU, Ankara, Turkey
	\end{itemize}
	
All the above mentioned groups are leaders in the emulsion scanning 
having gathered the experience of the emulsion analysis in  
 in the OPERA experiment. The scanning and the analysis of the exposed emulsions will be shared according to the available scanning power of each group. 

The development of the prototype, both for hardware and software, of the new read-out system is shared between LNGS and Naples while it is entirely carried out at Nagoya University for the Japanese one. 

The LNGS and Napoli groups are in charge of the design of the telescope, the construction of the prototype and  the calibration measurements. The same groups will perform the intrinsic background measurements, the studies about the environmental background and the design of the detector shielding and structure. 

The Russian groups will perform radioactive studies.

The design and the realization of the local underground facilities will be shared between LNGS and Japan.

The simulation of the detector response, efficiency and resolution as well as and the expected sensitivity is shared between LNGS, Naples and Nagoya.

The responsibility about the emulsion production, development and handling is currently assigned to the Nagoya group. 

\section{Conclusions and outlook} 

\begin{figure}[htbp]
	\centering
		\includegraphics[width=0.8\linewidth]{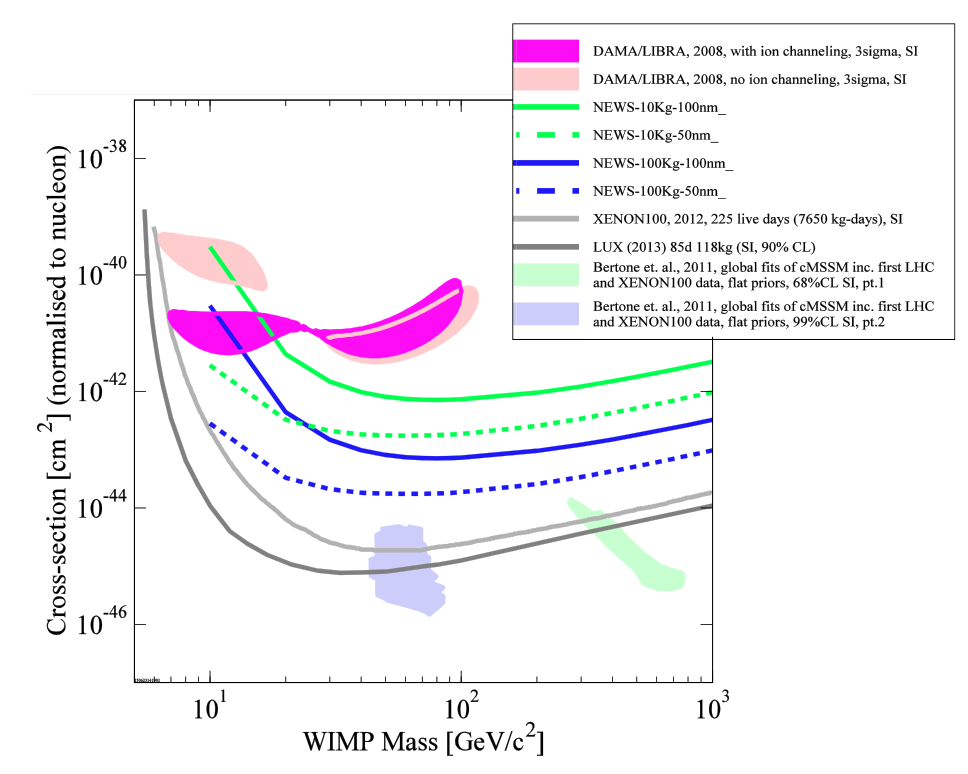}
	\caption{Sensitivity at 90$\%$ C.L, in the zero background hypothesis for an experiment with a mass of 10 kg (green) and 100 kg (blue) for two value of detection threshold: 100 nm (dashed lines) and 50 nm (solid line). }
	\label{fig:NEWSsensitivity_10-100Kg_50-100nm_JP}
\end{figure}

	NEWS is meant to be the first experiment with a solid target for directional dark matter searches: the use of a nuclear emulsion based detector, acting both as target and tracking device, would allow to explore the low cross section region in the phase space indicated by DAMA. 
	
	The novel emulsion technology, based on the use of nuclear emulsion with nanometric AgBr crystals (NIT), makes it possible to record the sub-micrometric tracks produced by the WIMP scattering  off a target nucleus. 
The presence, in the emulsion components, of light and heavy nuclei results in an enhanced sensitivity to both light and heavy WIMP masses.  
	
	The read-out of tracks with length of the order of 100 nm, is possible thanks to an R$\&$D carried out on the scanning systems currently used for the analysis of the OPERA emulsions. The use of improved optics and mechanics
	 allowed to reach a spatial and angular resolution of the order of 100 nm and 235 mrad, respectively, with a tracking efficiency approaching 100$\%$ for tracks with lengths longer than 180 nm. The new optical microscope has a scanning speed of about 25 mm$^2$/h allowing to perform a fast preselection of the candidate signal tracks with the shape analysis method. 
	
	The final signal confirmation is obtained with powerful optical microscope equipped with a light polarizer: exploiting the different response of non spherical grain clusters to different polarization angles, the unprecedented spatial resolution of 10 nm is obtained. This resolution allows to resolve grains belonging to a few hundred of nanometer long tracks thus providing the final signal confirmation with a  very high signal to noise ratio. 
	
	The intrinsic radioactivity of nuclear emulsions has been measured and a detailed MC simulation has been performed: the estimated neutron yield allows to design an experiment with 
	masses of the order of 10 kg while keeping this background negligible. 
 A careful evaluation of the external background sources has been performed allowing to design a proper shielding. The final experimental set-up foresees the use of an equatorial telescope holding both the emulsion target and the shielding. 
		
		We plan to perform a pilot experiment with a 1 kg mass target on a time scale of six years: even using a rather small detector mass we would be able to explore the region indicated by the DAMA experiment with a powerful and complementary approach (see Figure~\ref{fig:sensitivity1Kg}). 
	
	The actual intrinsic radioactive level allows to scale the target mass and exposure time up to one order of magnitude. A careful selection of the emulsion components and a better control of their production could further increase the radiopurity, thus allowing larger detector mass. The reduction of the fog density and further developments of the optical microscopy with polarized light would allow to reduce the detection threshold down to 50 nm. Improvements both in the mechanics (use of piezoelectric-driven objective) and in the image acquisition (use of multiple image sensors) envisage already now the possibility to analyse with such a resolution a volume of 100 kg or larger.  Moreover further improvements both in the microscope hardware and in the analysis software will permit to fully exploit the intrinsic emulsion capability of recording 3D tracks.

	In Figure~\ref{fig:NEWSsensitivity_10-100Kg_50-100nm_JP} the upper limit in case of null observation  for an experiment with a mass of 10 (green) and 100 (blue) kg and for a detection threshold of 50 (dashed lines) and 100 (solid lines) nm is shown at 90 $\%$ C.L and in the zero background hypothesis.
	
	The proposed program would open a new window in the DM search.
	The developments done will likely have impact on the nano-imagining applications in physics, biology and medicine. 
	
%\section*{Bibliography}

\end{document}